\documentclass[useAMS,usenatbib]{mn2e} 
\usepackage{graphics, graphicx} 

\newif\ifAMStwofonts

\def\lapp{\ifmmode\stackrel{<}{_{\sim}}\else$\stackrel{<}{_{\sim}}$\fi}
\def\gapp{\ifmmode\stackrel{>}{_{\sim}}\else$\stackrel{>}{_{\sim}}$\fi}

\title[High frequency observations of southern pulsars]
{High frequency observations of southern pulsars}
\author[Johnston, Karastergiou \& Willett]
{Simon Johnston$^1$, Aris Karastergiou$^2$ and Kyle Willett$^3$\\
$^1$Australia Telescope National Facility, CSIRO, P.O. Box 76, 
Epping, NSW 1710, Australia. \\
$^2$IRAM, 300 rue de la Piscine, Domaine Universitaire, Saint Martin
              d'H\`eres, France\\
$^3$Dept. of Physics and Astronomy, Carleton College, Northfield, MN 55057, USA.
}
\date{\today}
\pagerange{\pageref{firstpage}--\pageref{lastpage}}
\pubyear{2006}
\begin{document}
\maketitle
\label{firstpage}

\begin{abstract}
We present polarization data for 32 mainly southern pulsars at
8.4~GHz. The observations show that the polarization fraction is low
in most pulsars at this frequency except for the young, energetic pulsars
which continue to show polarization fractions in excess of 60 per cent.
All the pulsars in the sample
show evidence for conal emission with only one third also showing core
emission. Many profiles are asymmetric, with either the leading or the
trailing part of cone not detectable. Somewhat surprisingly, the
asymmetric profiles tend to be more polarized than the symmetrical
profiles. Little or no pulse narrowing is seen between 1 and
8.4~GHz. The spectral behaviour of the orthogonal polarization modes
and radius to frequency mapping can likely account for much of the
observational phenomenology. Highly polarized components may orginate
from higher in the magnetosphere than unpolarized components.
\end{abstract}


\section{Introduction}
It is generally accepted that low frequency radio emission
from pulsars arises from
further out the magnetosphere than high frequency
emission. The fact that the open magnetic field lines diverge outwards at
higher altitudes then implies that low frequency profiles are
generally broader than those at high frequency. However, this
broadening effect is most prominent at frequencies below about
1~GHz. Above this frequency, the profile narrows only slowly (if at
all) as pointed out by von Hoensbroech \& Xilouris (1997)\nocite{hx97a}
in their high frequency study.
Both Mitra \& Rankin (2001)\nocite{mr01} and Gangadhara \&
Gupta (2003)\nocite{gg03} showed that emission appears not
to occur throughout the polar cap but is concentrated in the inner
$\sim$60 per cent, again limiting the amount of observational widening
seen.

Observationally, individual components within a pulse profile can be
classified as `core' or `cone' components (Rankin 1983; Lyne \&
Manchester 1988; Rankin 1990)\nocite{ran83,lm88,ran90}.  Core emission
arises from near the magnetic axis whereas cone emission originates
from more distant regions of the magnetosphere. There appears to be a
difference in the spectral index of core and cone emission with core
emission having the steeper index. Theoretical considerations show
that these differences are perhaps caused by refraction in the
magnetosphere, where the lower plasma density in the central (core)
regions cause enhanced refraction and effectively a steeper spectral
index \citep{pet02}. The observational implications are that low
frequency profiles can often be dominated by core emission whereas at
high frequencies, prominent conal outriders can often be detected. 
However, it is not clear cut that two different emission mechanisms
operate. Kramer et al. (1994)\nocite{kwj+94} and 
Sieber (1997)\nocite{sie97} showed that a number of the
observational effects could be attributed to geometric arguments and
Lyne \& Manchester (1988) prefer a patchy beam model.
von Hoensbroech et al. (1998)\nocite{hkk98} also showed that there
exists a `class' of pulsars which has one highly polarized component
usually on the extreme leading or trailing edge of the profile.
The highly polarized component has a flat spectral index and is therefore
prominent at high frequencies. The presence of dominant outrider
components at high frequencies compared to low frequencies can
actually {\it increase} the pulse width and some care must be taken
when comparing results at different frequencies.

In general, the fractional linear polarization decreases with
increasing frequency. Competition between orthogonal polarization
modes has been shown to result in such an effect
\citep[e.g.][]{kkj+02}. The steep decrease in fractional linear
polarization seen in some pulsars at very high frequencies (above
$\sim$10~GHz) may warrant an alternative explanation
\citep{xkj+96}. In any case, young pulsars form a special group by
retaining their high fractional polarization over a wide frequency
range. In some pulsars, the high linear polarization seen at low
frequencies decreases, while at the same time the circular
polarization increases. These are observational examples of the
`conversion' of linear to circular polarization discussed by
\cite{hl99}. The position angle swing is generally expected to have
its largest slope at the longitude at which the line of sight crosses
the magnetic meridian. Core components therefore tend to have steep PA
swings across them, whereas conal components show only shallow PA
swings. Core components sometimes also show a swing in the sign of the
circular polarization; this is rarely observed in conal components.

It is worth noting that single-pulse studies reveal a different side
of pulsar emission relating to short timescale changes in the pulsar
magnetosphere. Integrated profiles and single pulses often show
important observational differences. For example, the spectral index
measured in single pulses does not necessarily show the same
dependence on the type of component \citep{kkg+03} as discussed
above. Also, in the single pulses a variety of polarization phenomena
are observed, such as swings in the sign of the circular polarization
in cone components \citep{khk+01}, which disappear in the averaging
process \citep{kjm+03}.

This paper is the fourth in a series of recent papers examining high
frequency polarization of integrated profiles of southern pulsars.  In
the first paper Karastergiou, Johnston \& Manchester (2005; hereafter
KJM05\nocite{kjm05}) observed 48 pulsars at 3.1~GHz and considered the
polarization evolution with frequency in the context of competing
orthogonal modes. They developed an empirical model for the frequency
evolution of the linear polarization based on different spectral
indices of the orthogonally polarized modes. 
In the second paper, Karastergiou \& Johnston (2006; hereafter
KJ06\nocite{kj06}) concentrated on 17 strong pulsars at both 1.4 and
3.1~GHz and made a careful comparison of the position angle (PA) swing
at the two frequencies. Finally, Johnston \& Weisberg
(2006)\nocite{jw06} looked at the morphology and polarization of a
group of 14 young pulsars and showed that emission arises from high in
the magnetosphere with little or no core emission and highly polarized
conal emission.

In this paper we examine a group of the strongest pulsars at 8.4~GHz
in order to study the total intensity profile and polarization. We
use previous observations of these pulsars made at lower frequencies 
to draw conclusions
on the frequency evolution of these properties. Apart from a small
number of isolated exceptions, these are the first observations of
these pulsars above 3~GHz and the first systematic study of southern
pulsars at high frequencies.

\section{Observations}
\begin{table*}
\caption{The observed pulsars and their parameters. Numbers in brackets
denote the value for the interpulses. Values for percentage linear (\%L)
and circular (\%L) refer to the entire profile; individual components
can differ significantly from this.}
\begin{tabular}{llrrrrrr}
\hline & \vspace{-3mm} \\
Jname & Bname & Period & Age   & S$_{8.4}$ & W$_{10}$ & \%L & \%V \\
      &       & (ms)   & (Myr) & (mJy)     &   (deg)  \\
\hline & \vspace{-3mm} \\
J0630$-$2834 & B0628$-$26 & 1244.4 & 2.77 & 2.3 & 20 & $<$7 & $<$7\\
J0659+1414   & B0656+14   & 384.9 & 0.11 & 2.5 & 26 & $<$7 & $<$7\\
J0738$-$4042 & B0736$-$40 & 374.9 & 3.7 & 7.0 & 24 & 11$\pm$2 & 3$\pm$2\\
J0742$-$2822 & B0738$-$28 & 166.8 & 0.16 & 1.6 & 14 & 57$\pm$4 & $-14\pm$4\\
J0835$-$4510 & B0833$-$45 & 89.3 & 0.011 & 4.1 & 18 & 80$\pm$1 & $-27\pm$1\\
J0837$-$4135 & B0835$-$41 & 751.6 & 3.4 & 1.3 & 13 & 15$\pm$3 & $-6\pm$3\\
J0908$-$4913 & B0906$-$49 & 106.8 & 0.11 & 0.4(1.2) & 15(10) & 75$\pm$4(71$\pm$1) & $<$4($-15\pm$1)\\
J0922+0638   & B0919+06   & 430.6 & 0.50 & 0.6 & 7 & 15$\pm$12 & $<$12\\
J0953+0755   & B0950+08   & 253.1 & 17.5 & 4.3 & 30 & $<$4 & $<$4\\
J1048$-$5832 & B1046$-$58 & 123.7 & 0.020 & 3.2 & 19 & 78$\pm$1 & 19$\pm$1\\
J1056$-$6258 & B1054$-$62 & 422.4 & 1.9 & 5.1 & 30 & 10$\pm$2 & $<$2\\
J1136+1551   & B1133+16   & 1187.9 & 5.0 & 0.6 & 8 & 4$\pm$4 & $<$4\\
J1243$-$6423 & B1240$-$64 & 388.5 & 1.4 & 2.0 & 14 & $<$15 & $<$15\\
J1302$-$6350 & B1259$-$63 & 47.7 & 0.33 & 0.9(1.4) & 20(30) & 60$\pm$8(66$\pm$7) & 15$\pm$8($<$7)\\
J1326$-$5859 & B1323$-$58 & 478.0 & 2.3 & 1.3 & 17 & $<$8 & $<$8\\
J1327$-$6222 & B1324$-$62 & 529.9 & 0.44 &     & 12 & 14$\pm$3 & 7$\pm$3\\
J1341$-$6220 & B1338$-$62 & 193.3 & 0.012 & 0.8 & 12 & 67$\pm$4 & 6$\pm$4\\
J1359$-$6038 & B1356$-$60 & 127.5 & 0.32 & 2.2 & 13 & 39$\pm$8 & 20$\pm$8\\
J1430$-$6623 & B1426$-$66 & 785.4 & 4.5 & 0.6 & 7 & 11$\pm$11 & $<$11\\
J1453$-$6413 & B1449$-$64 & 179.5 & 1.0 & 1.5 & 10 & 19$\pm$10 & $<$10\\
J1456$-$6843 & B1451$-$68 & 263.4 & 42.5 & 3.4 & 24 & $<$3 & $<$3\\
J1522$-$5829 & B1518$-$58 & 395.4 & 3.1 & 1.4 & 15 & 21$\pm$8 & 20$\pm$8\\
J1539$-$5626 & B1353$-$56 & 243.4 & 0.80 & 2.3 & 23 & 39$\pm$7 & 10$\pm$7\\
J1600$-$5044 & B1557$-$50 & 192.6 & 0.60 & 1.9 & 8 & 18$\pm$9 & $<$9\\
J1602$-$5100 & B1558$-$50 & 864.2 & 0.20 & 0.9 & 4 & $<$11 & $<$11\\
J1630$-$4733 & B1627$-$47 & 576.0 & 0.41 & 2.9 & 20 & 18$\pm$5 & $-10\pm$5\\
J1644$-$4559 & B1641$-$45 & 455.1 & 0.36 & 2.2 & 16 & 39$\pm$1 & $-4\pm$1\\
J1709$-$4429 & B1706$-$44 & 102.5 & 0.018 & 5.5 & 32 & 72$\pm$3 & $-15\pm$1\\
J1721$-$3532 & B1718$-$35 & 280.4 & 0.18 & 1.9 & 17 & 25$\pm$1 & $-9\pm$1\\
J1730$-$3350 & B1727$-$33 & 139.5 & 0.026 & 0.8 & 12 & 88$\pm$16 & $-35\pm$16\\
J1740$-$3015 & B1737$-$30 & 606.7 & 0.021 & 0.5 & 8 & 48$\pm$4 & $-60\pm$4\\
J1752$-$2806 & B1749$-$28 & 562.6 & 1.1 & 0.9 & 12 & $<$10 & $<$10\\
\hline & \vspace{-3mm} \\
\end{tabular}
\end{table*}

Observations were carried out using the 64-m radio telescope located
near Parkes, New South Wales. The central frequency was 8.356~GHz with
a total bandwidth of 512 MHz. The feed consists of dual circular
receptors and a calibrator probe at 45 degrees to the feed probes.
During the observing run, observations were made of the flux
calibrator Hydra~A whose flux density is 8.5~Jy at this frequency.
This allowed us to determine the system equivalent flux density to be
48~Jy.

The pulsars were typically observed for 30~mins each, preceded by a
2~min observation of a pulsed calibrator signal. The total bandwidth
was subdivided into 1024 frequency channels and the pulsar period
divided into 1024 phase bins by the backend correlator. The correlator
folds the data for 60~s at the topocentric period of the pulsar at
that epoch and records the data to disk for offline-processing.

The data were taken in two periods. The first ran from 2005 May 4 to
2005 May 10. Unfortunately many of the resultant pulse profiles were 
ruined by systematic
problems in the backend system. Although the total intensity and
circular polarization profiles were generally reasonable the Stokes $Q$
and $U$ profiles were completely corrupted.
However, the data were useful in that we were able to determine which pulsars
showed good signal to noise in 30~min and which were undetectable.
We obtained additional time between 2005 July 8 and July 12 and 
re-observed 32 pulsars for which we had convincing detections from 
the earlier session.

The data were analysed off-line using the PSRCHIVE software package
\citep{hvm04}. Polarization calibration was carried out using
the observations of the pulsed calibrator signal to determine the relative
gain and phase between the two feed probes. The data were corrected
for parallactic angle and the orientation of the feed. The position
angles were also corrected for Faraday rotation through the interstellar
medium using the nominal rotation measure (at this high frequency
errors in the rotation measure make very little impact on the absolute
position angles). Flux calibration
was carried out using the Hydra~A observations. The final product was
therefore flux and polarization calibrated Stokes $I$, $Q$, $U$, $V$
profiles.

\section{Categorising the profiles}
The classification of pulsars into categories based on their
observational properties is a natural step in attempting to understand
the underlying physics. Rankin (1983) is a pioneer in this field and
her subsequent papers lay out a series of ground rules for the
classification of pulse profiles at frequencies around $\sim$1~GHz.
However, not only are useful criteria for the classification difficult
to decide upon, it is also often hard to unambiguously identify
individual components within a complex pulse morphology, especially
when the components have different frequency dependent behaviour. In
this section we outline a scheme for classifying pulsars at this high
frequency of 8.4~GHz.

By and large, virtually all pulsars at 8.4~GHz show conal emission of
some sort and we should not expect any core only pulsars. This is not
overly surprising. Core dominated pulsars are likely to be steep
spectrum objects and therefore not above our detection threshold
limit.  Secondly conal emission has a flatter spectral index and is
more likely to be seen at higher frequencies.  We therefore introduce
two broad categories. The first contains pulsars in which only conal
emission is present and the second contains pulsars for which both
core and cone emission can be seen. Each of these cateogories has
a number of sub-classes based on the degree of symmetry of the
profile.

\subsection {Profiles without core emission}
\subsubsection {Symmetric cone distribution}
Profiles in this category are those with no evident central component
at any frequency and with a symmetric (and perhaps multiple)
distribution of cones. The polarization tends to be low and decreases
further with increasing frequency. The pulse profile should also
narrow with increasing frequency as expected.  In some profiles the
components are blended together whereas in others the two components
are clearly detached.  The spectral index of the individual components
can be varied with either leading or trailing components dominating
the profile.

\subsubsection {Asymmetric cone distribution}
The pulsars in this category show only partial (one-sided) conal
emission and no obvious core. There are two sub-types, those showing
leading edge emission, generally characterised by profiles with a
steep rising edge and a slower falling (inner) edge and those showing
trailing edge emission where the reverse is true. Generally, rather
little profile evolution is seen and there is no narrowing of the
pulse width. The PA variation tends to be flat and constant over the
pulse. Finally, it may be that there is emission along the magnetic
meridian but which is distant (in latitude) from the pole itself.  We
consider this as grazing conal emission, often seen in young pulsars
and characterised by a constant, but significant PA swing.

\subsection {Profiles with core emission}
\subsubsection {Symmetric cone distribution}
The pulsars in this category show `classic'
behaviour as a function of frequency. At low frequencies the profiles
are dominated by a central component whereas at high frequencies the
outrider components are well separated from the central component and
dominate the profile to a greater or lesser extent.

\subsubsection {Asymmetric cone distribution}
In this category there is evidence for both core and
cone emission in the pulse profiles with only one side of the
cone visible (either trailing edge of leading edge).
These pulsars show significant frequency evolution with
the core reducing in prominence and the cone increasing in prominence
with increasing frequency. Generally a flat PA swing is seen across
the cone component with a steep swing of PA across the core component. 

\subsection{Young pulsars}
Young pulsars appear to form their own class of pulse profiles
\citep{qmlg95,man96}.  The morphology and polarization
of young pulsars in general have most recently been discussed 
in Johnston \& Weisberg (2006). Their profiles are consistent with being 
conal doubles, or grazing edge cones, with little or no core emission.
In virtually all cases the trailing edge of the cone dominates the profile.
There is very little profile evolution as a function of frequency and
the total polarization fraction remains very high at all frequencies.
von Hoensbroech \& Lesch (1999)\nocite{hl99} showed that some pulsars 
in this category appear to `convert' linear to circular polarization 
at high frequencies.

\section{Individual pulse profiles}
This section details the polarization profiles of the 32 pulsars in
our sample. Table~1 lists the pulsars along with their spin periods
and age. Columns 5 and 6 of the table give the flux density at
8.4~GHz and the width of the profile at the 10 per cent level.
The final two columns list the percentage linear and circular averaged
over the profile. Note that the percentage polarization in individual 
components within the profile can vary significantly. 
Descriptions of the individual pulsars are given below and their
polarization profiles shown in Figs 1 to 4.
\begin{figure*}
\begin{center}
\begin{tabular}{cc}
\resizebox{0.4\hsize}{!}{\includegraphics[angle=-90]{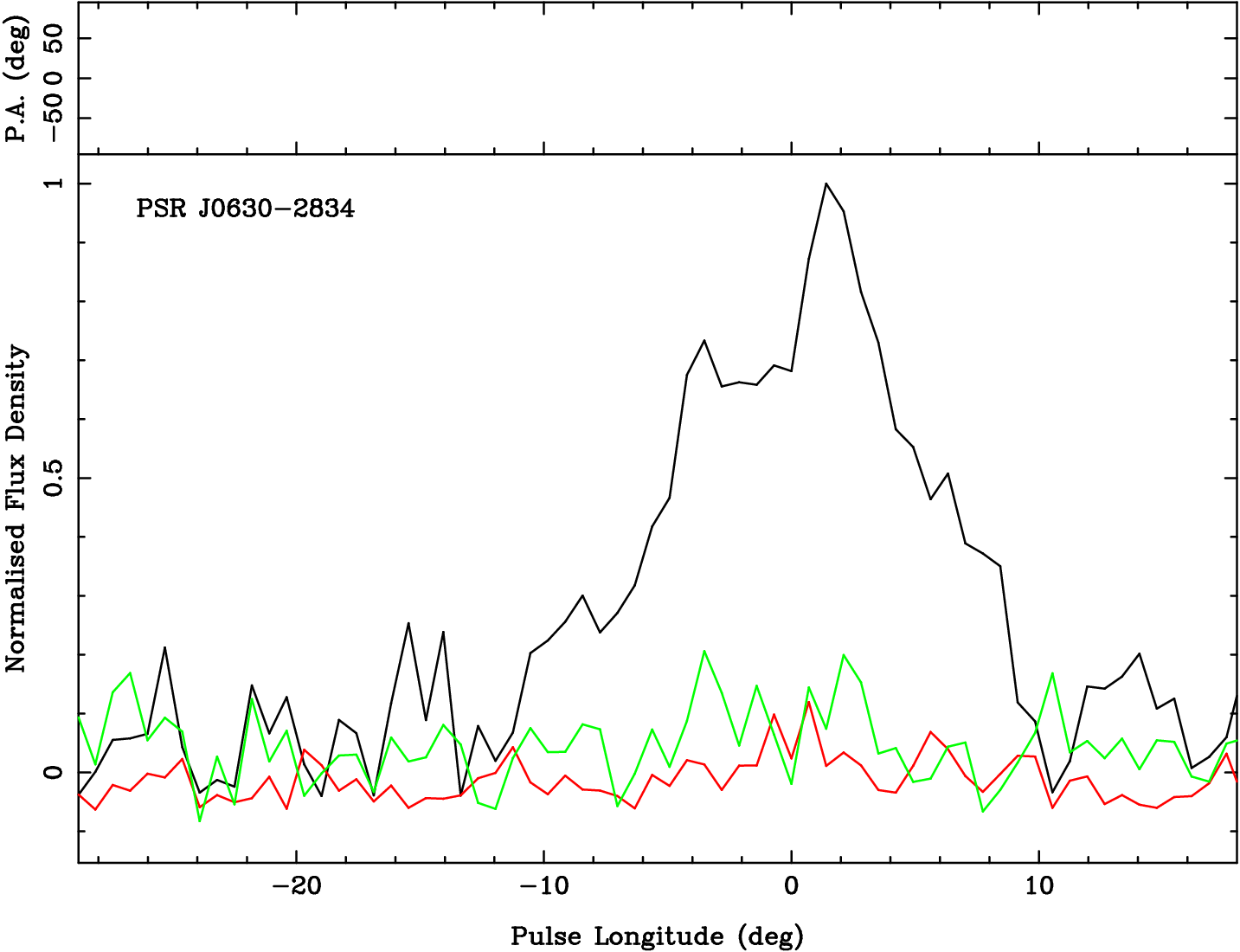}}&
\resizebox{0.4\hsize}{!}{\includegraphics[angle=-90]{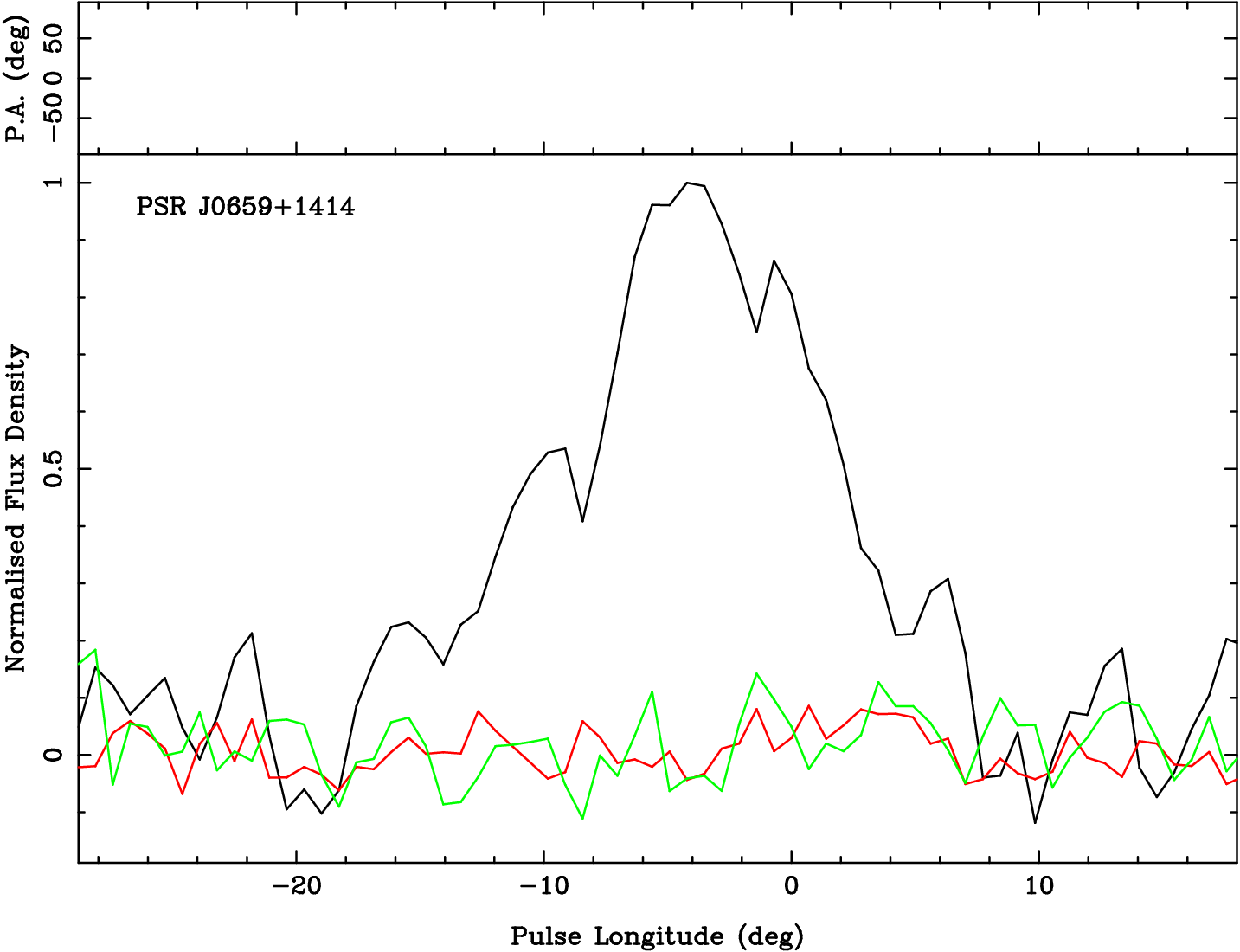}}\\
\resizebox{0.4\hsize}{!}{\includegraphics[angle=-90]{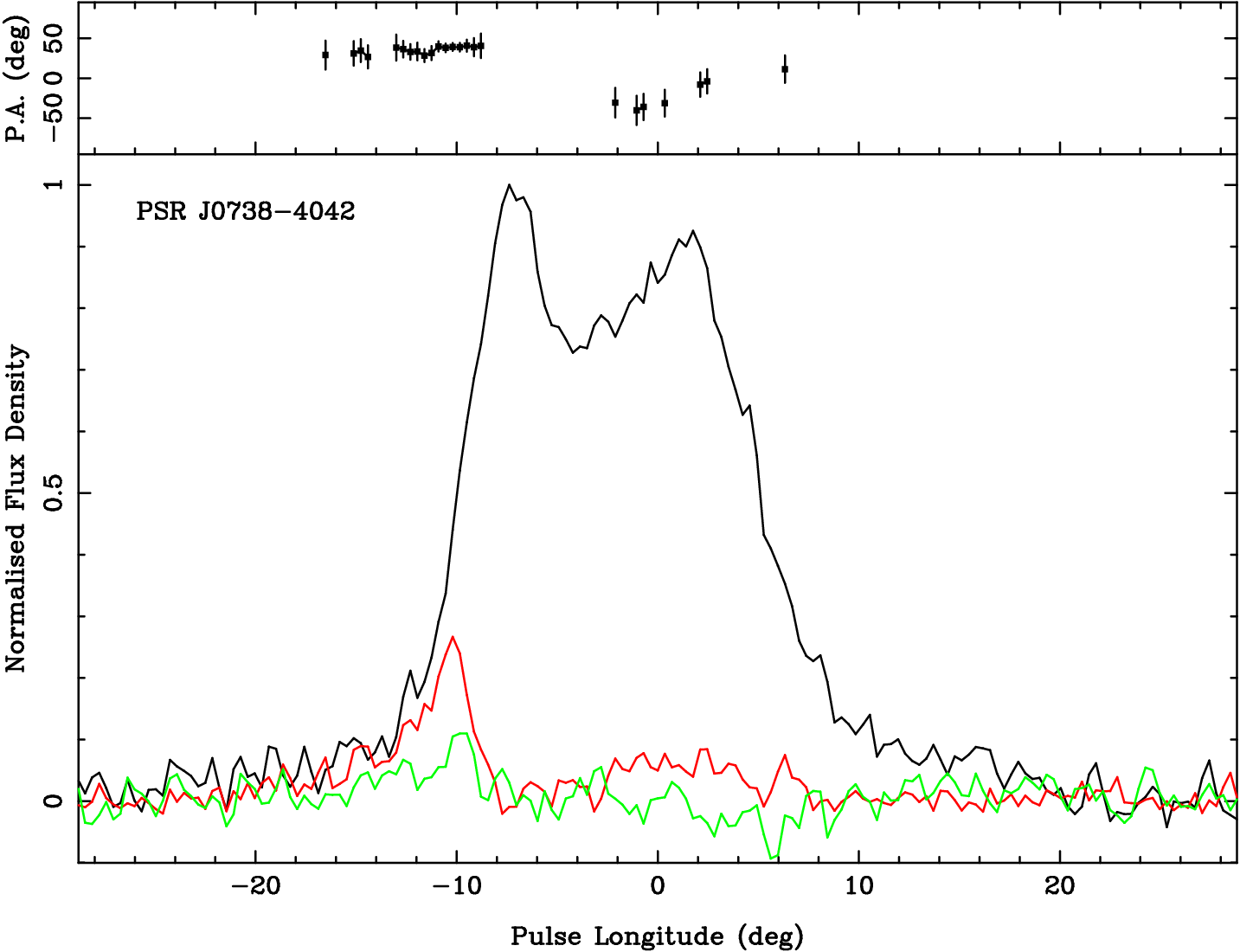}}&
\resizebox{0.4\hsize}{!}{\includegraphics[angle=-90]{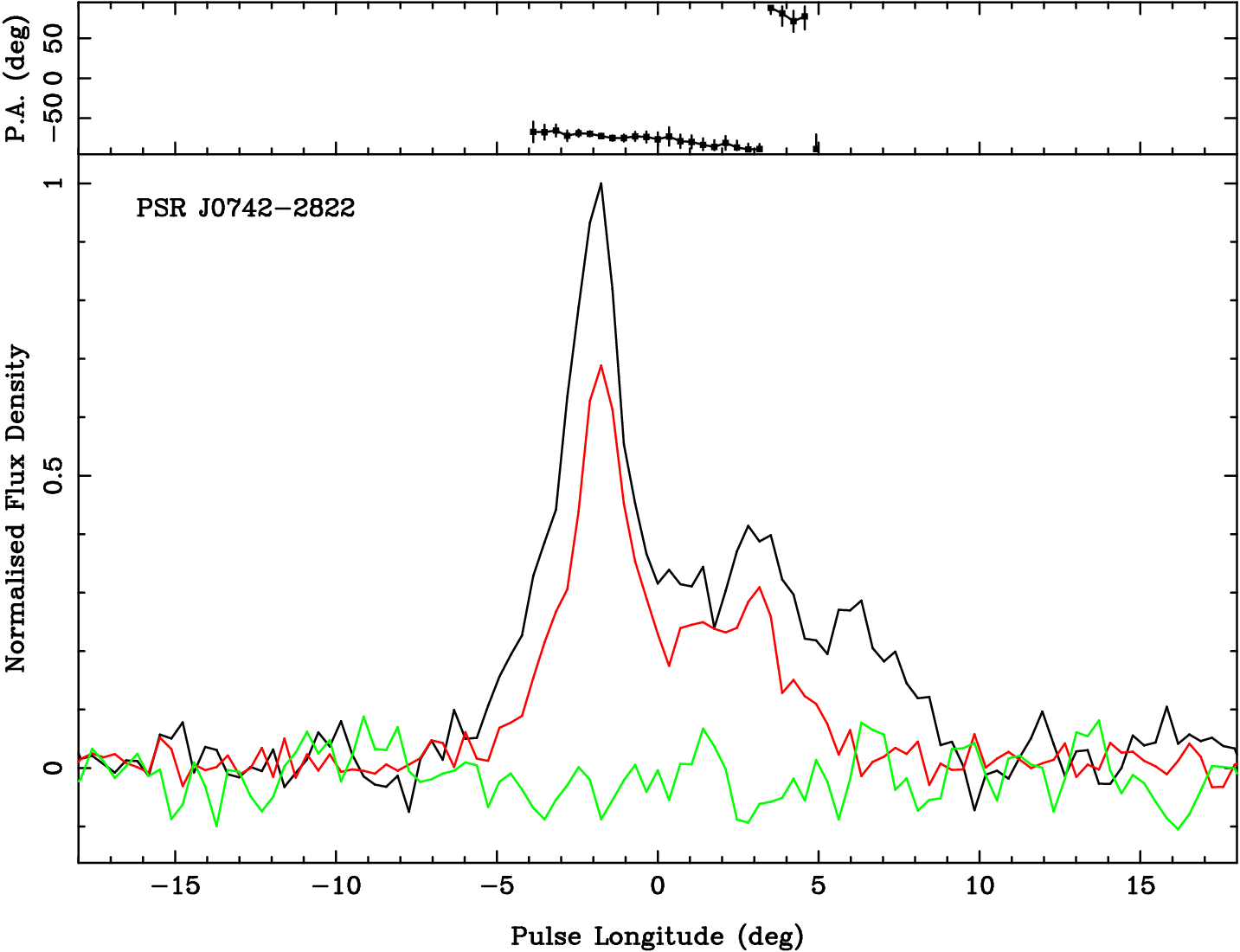}}\\
\resizebox{0.4\hsize}{!}{\includegraphics[angle=-90]{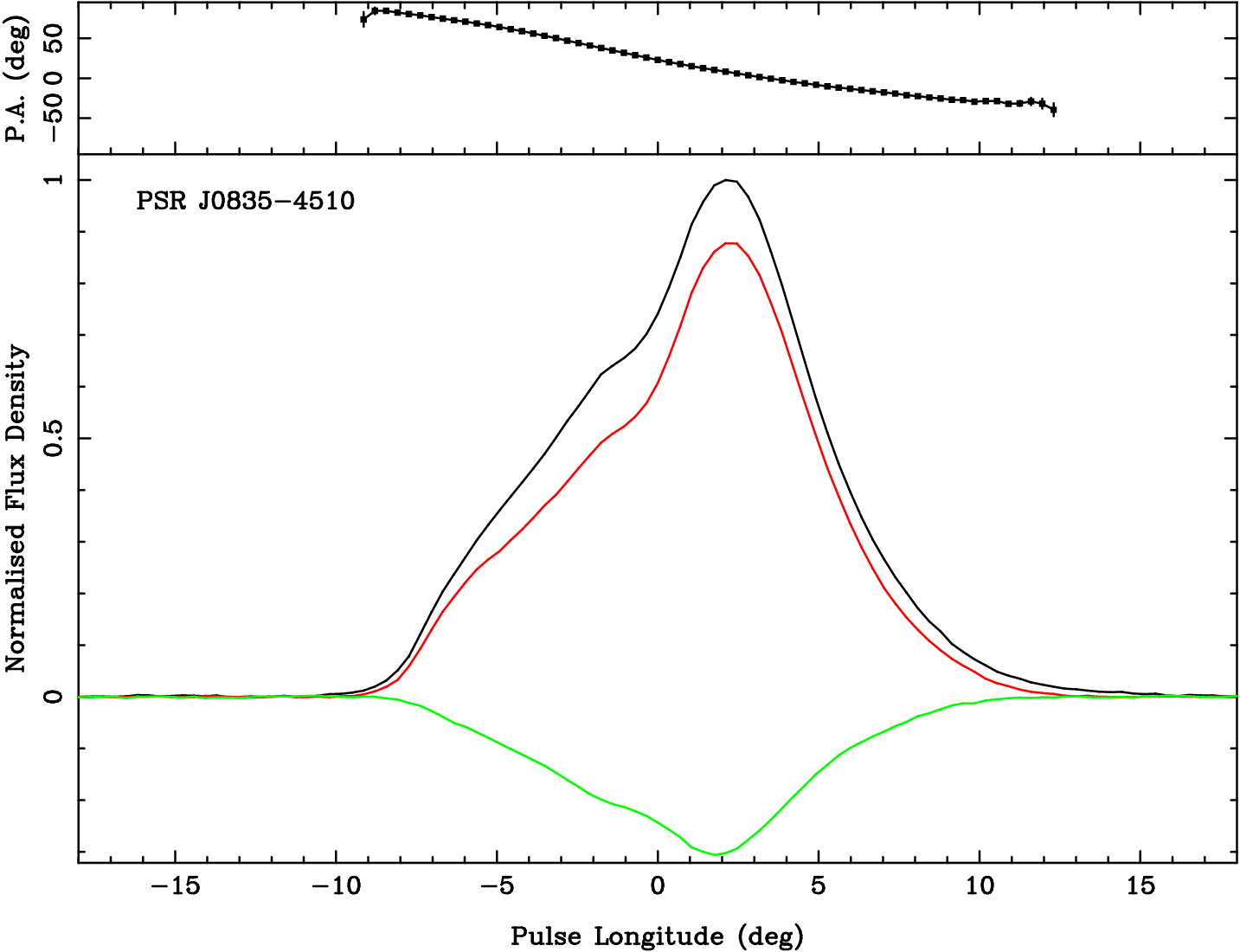}}&
\resizebox{0.4\hsize}{!}{\includegraphics[angle=-90]{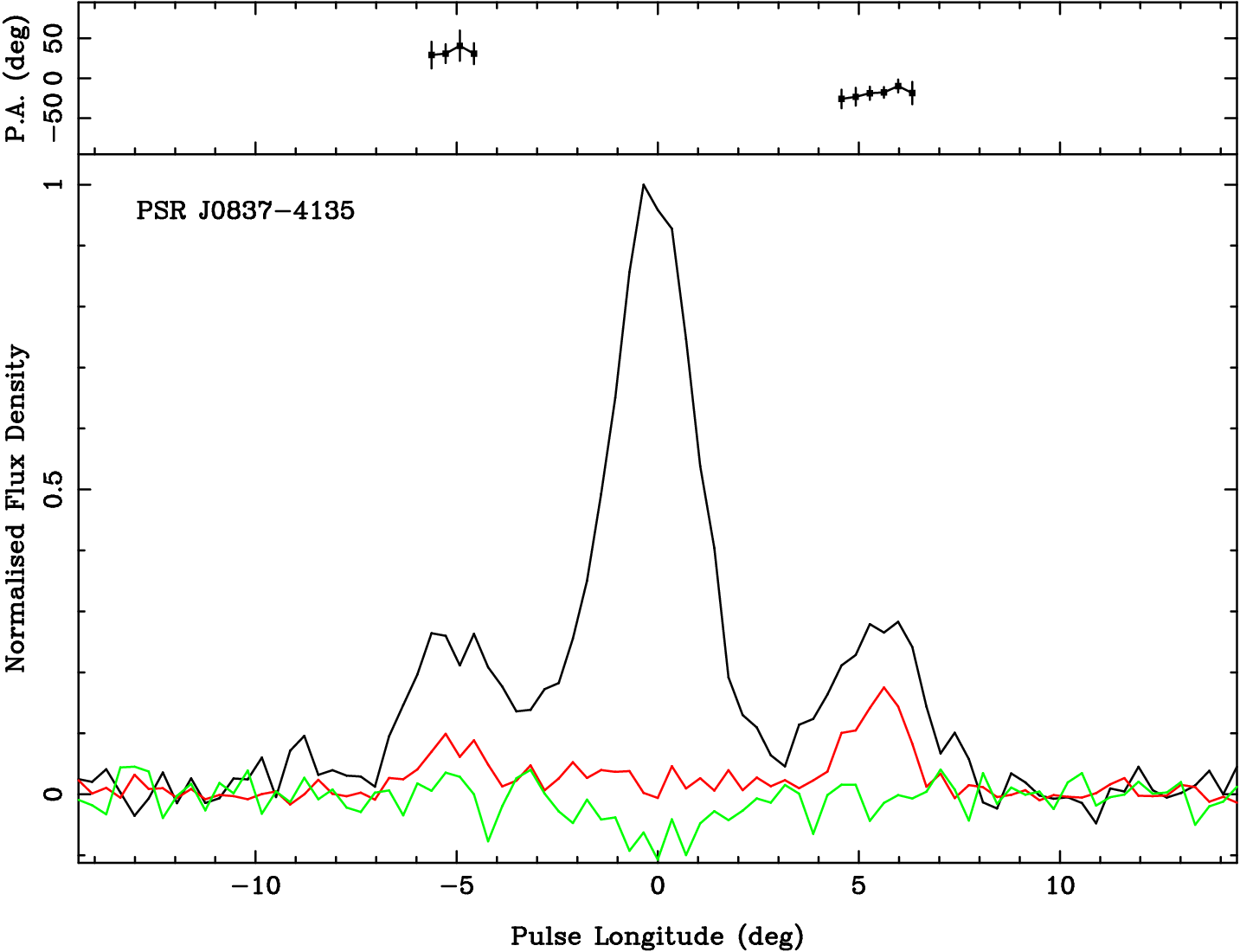}}\\
\resizebox{0.4\hsize}{!}{\includegraphics[angle=-90]{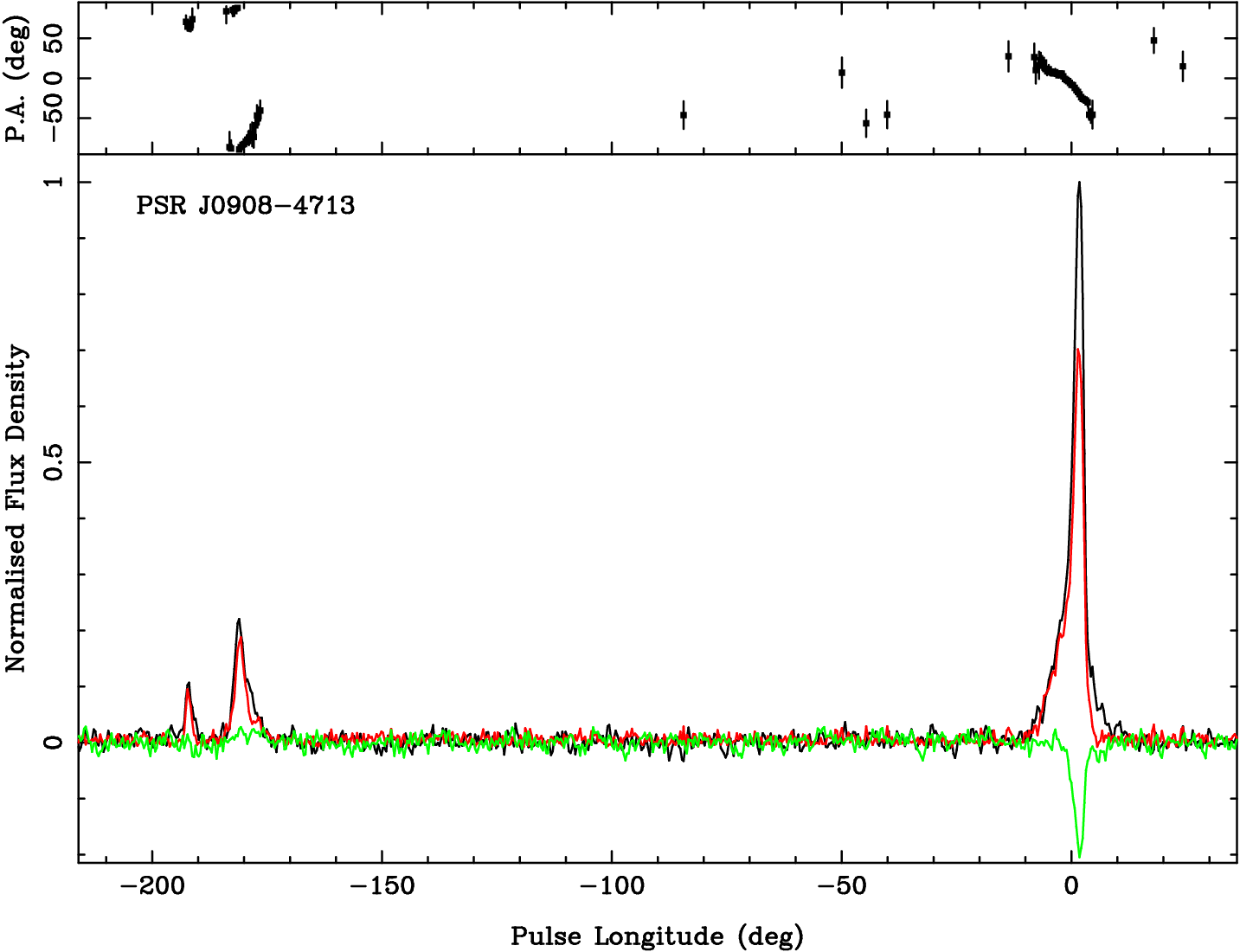}}&
\resizebox{0.4\hsize}{!}{\includegraphics[angle=-90]{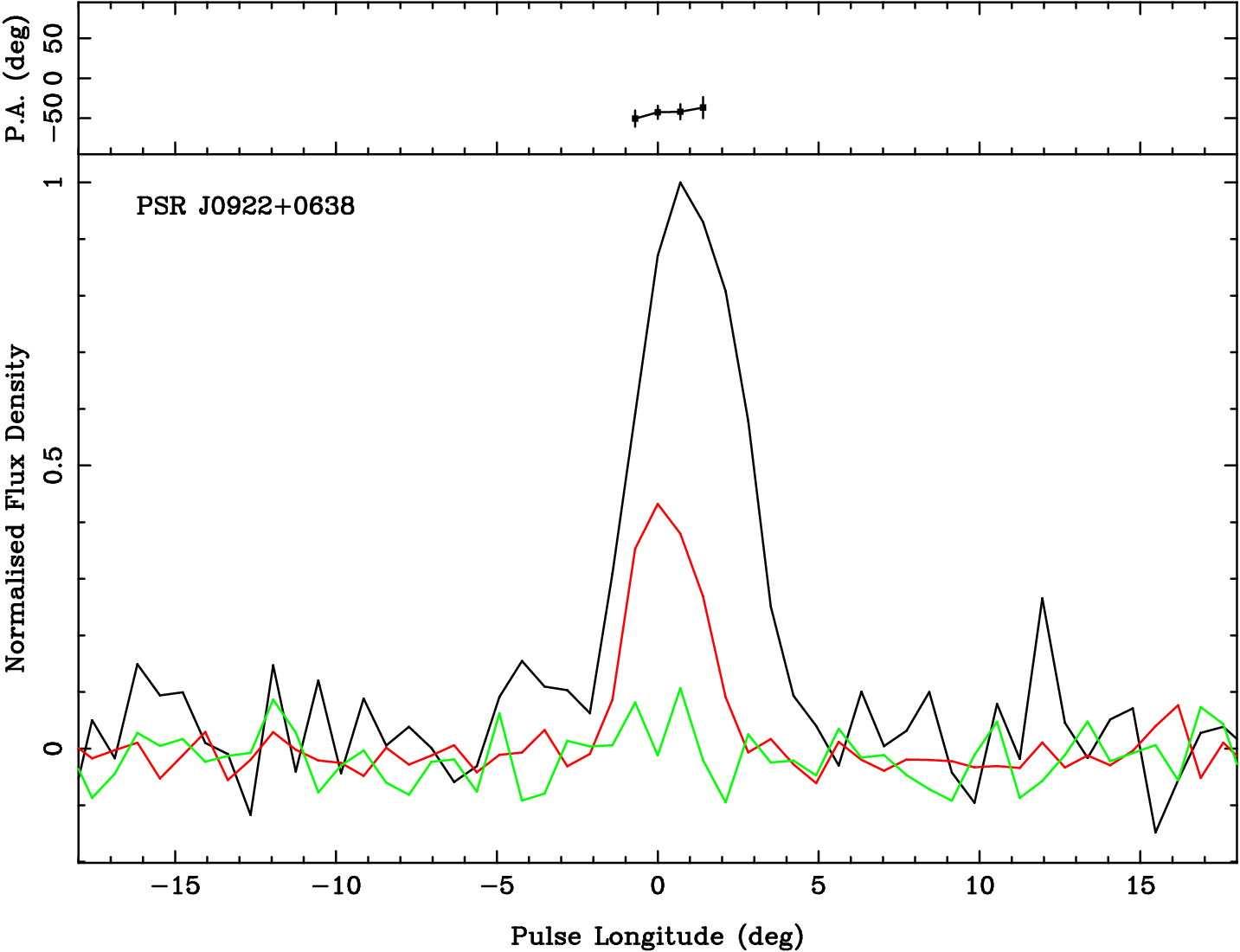}}\\
\end{tabular}
\end{center}
\caption{Polarisation profiles at 8.4~GHz for 8 pulsars as marked.
The top panel of each plot shows the PA variation with respect to
celestial north as a function of longitude.  The PAs are corrected for
RM and represent the (frequency independent) value at the pulsar.  The
lower panel shows the integrated profile in total intensity (thick
line), linear polarization (dark grey line) and circular polarization
(light grey line).}
\end{figure*}
\begin{figure*}
\begin{center}
\begin{tabular}{cc}
\resizebox{0.4\hsize}{!}{\includegraphics[angle=-90]{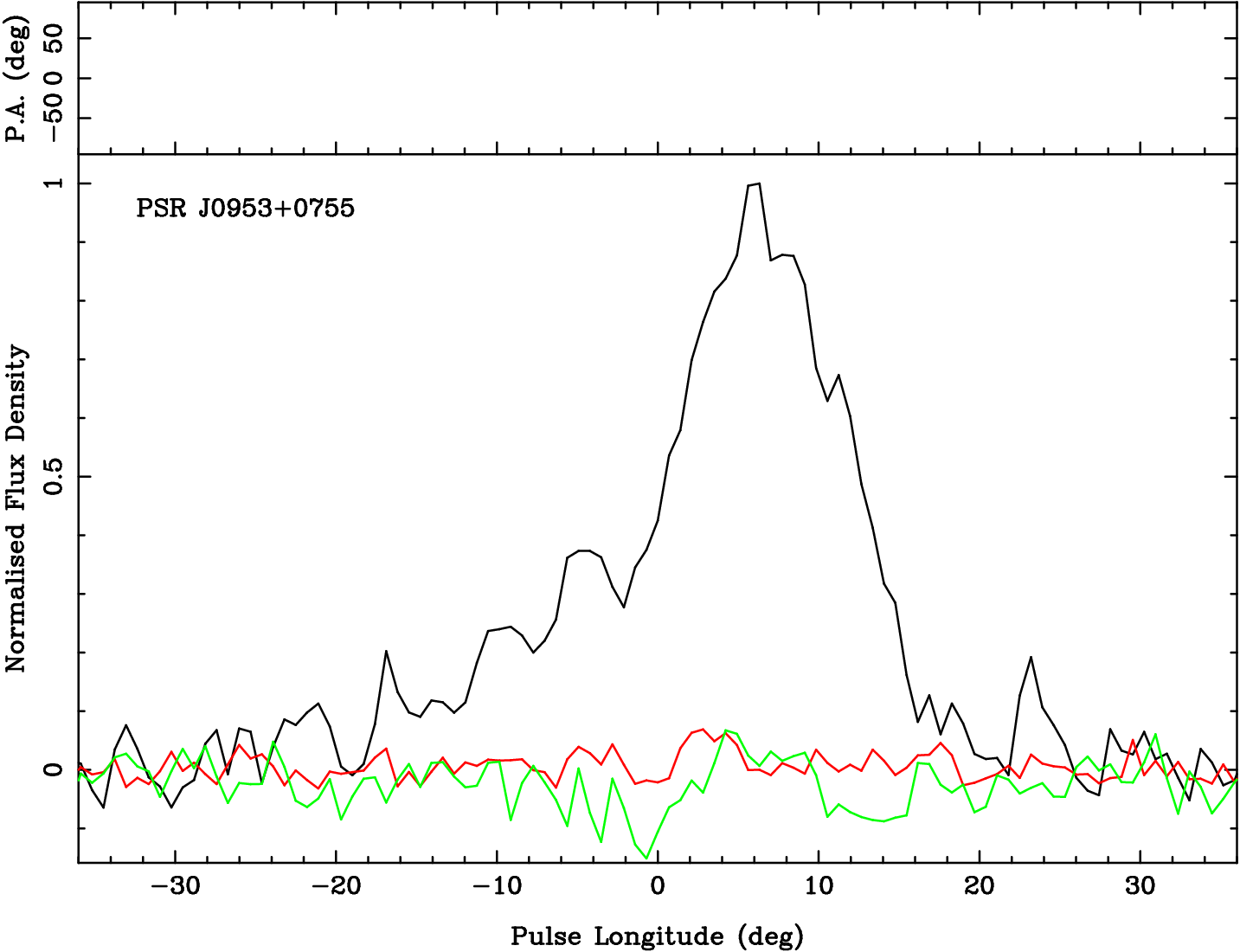}}&
\resizebox{0.4\hsize}{!}{\includegraphics[angle=-90]{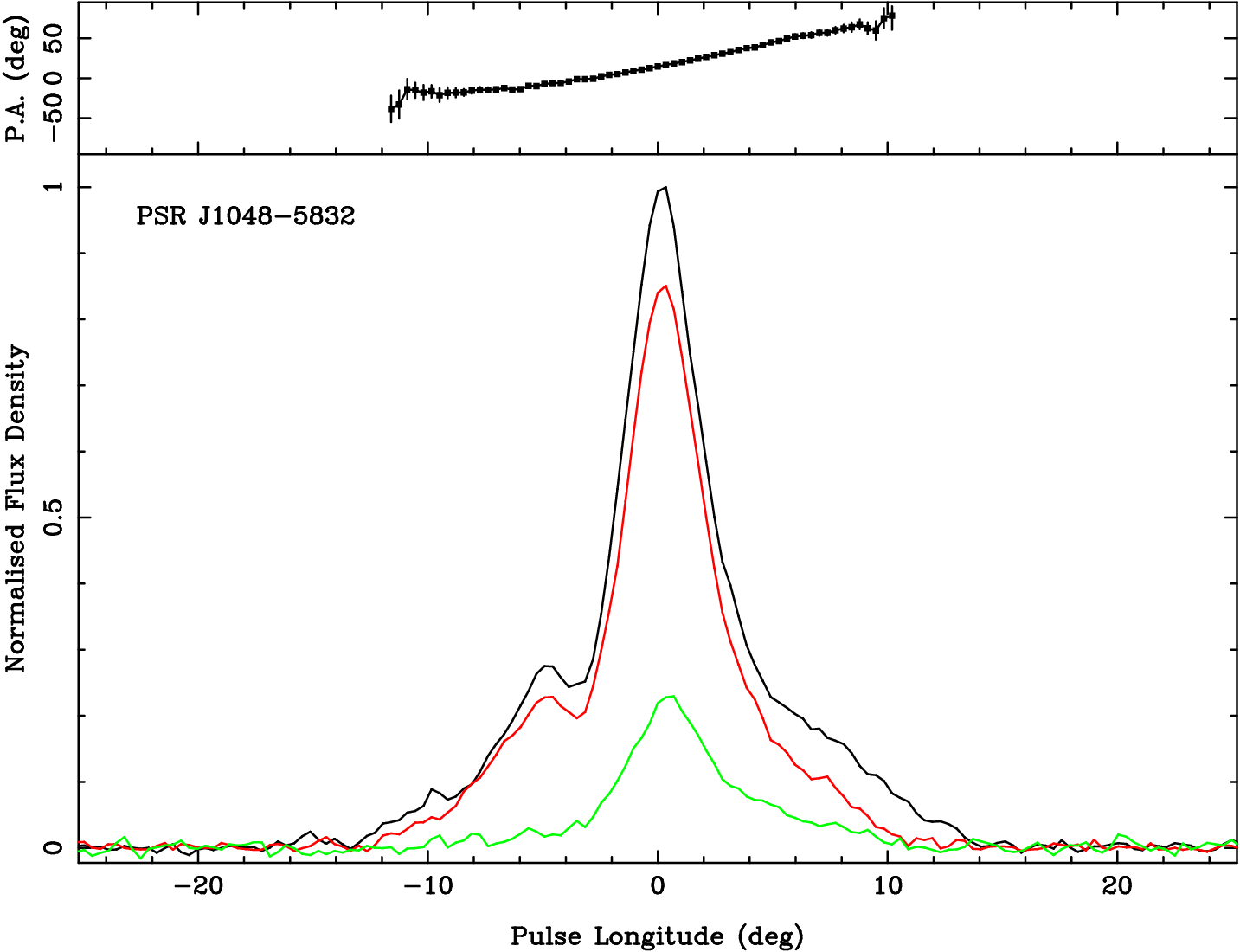}}\\
\resizebox{0.4\hsize}{!}{\includegraphics[angle=-90]{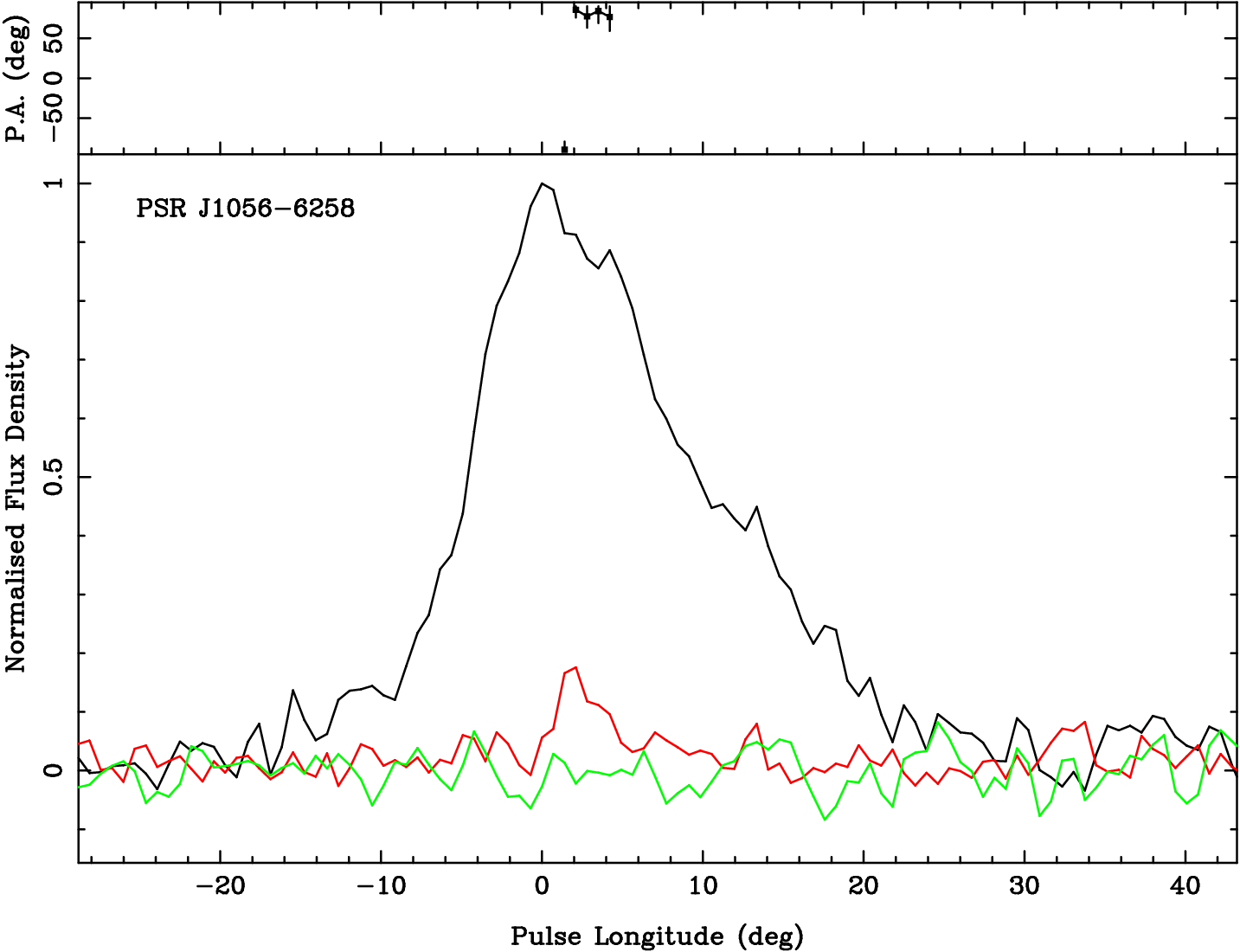}}&
\resizebox{0.4\hsize}{!}{\includegraphics[angle=-90]{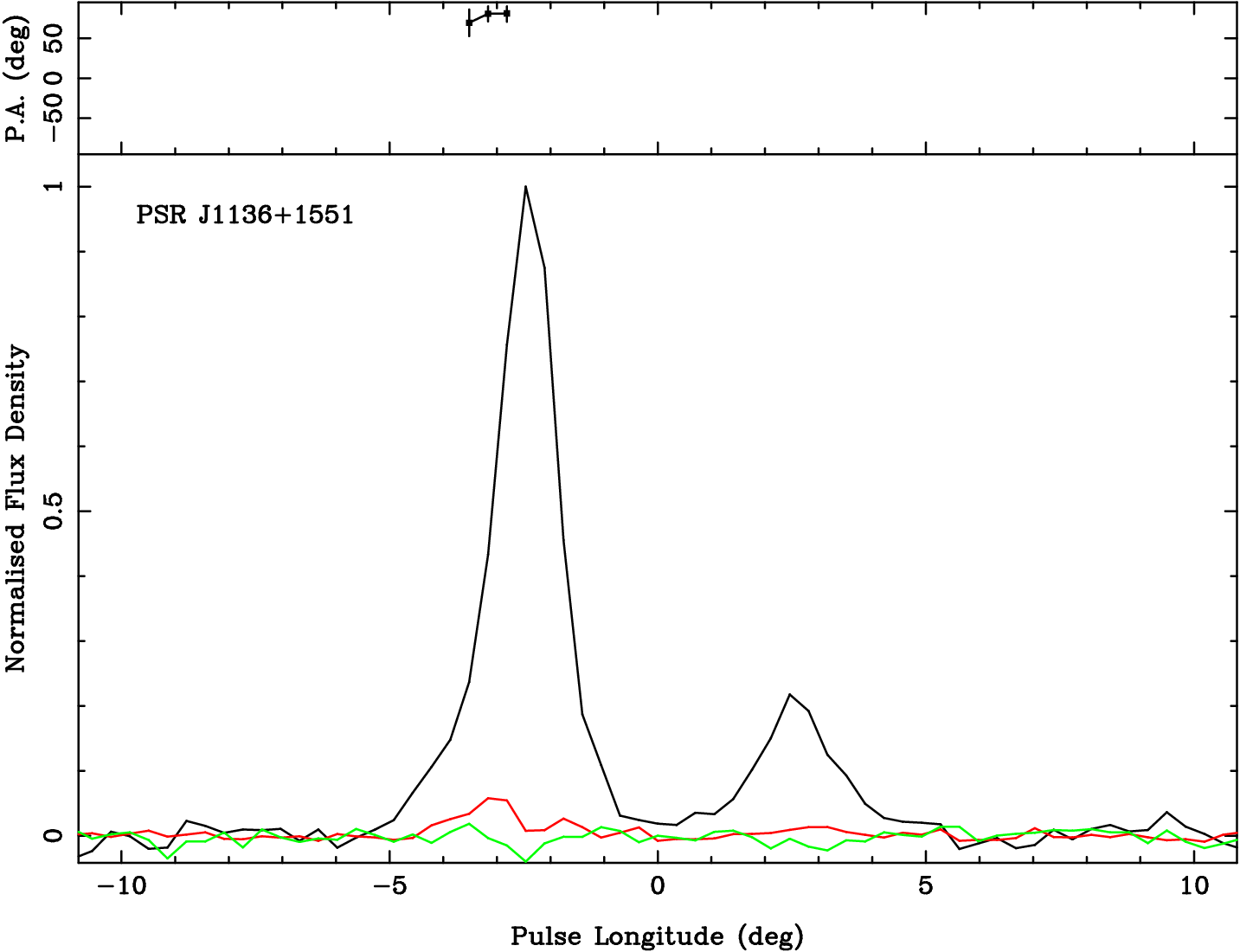}}\\
\resizebox{0.4\hsize}{!}{\includegraphics[angle=-90]{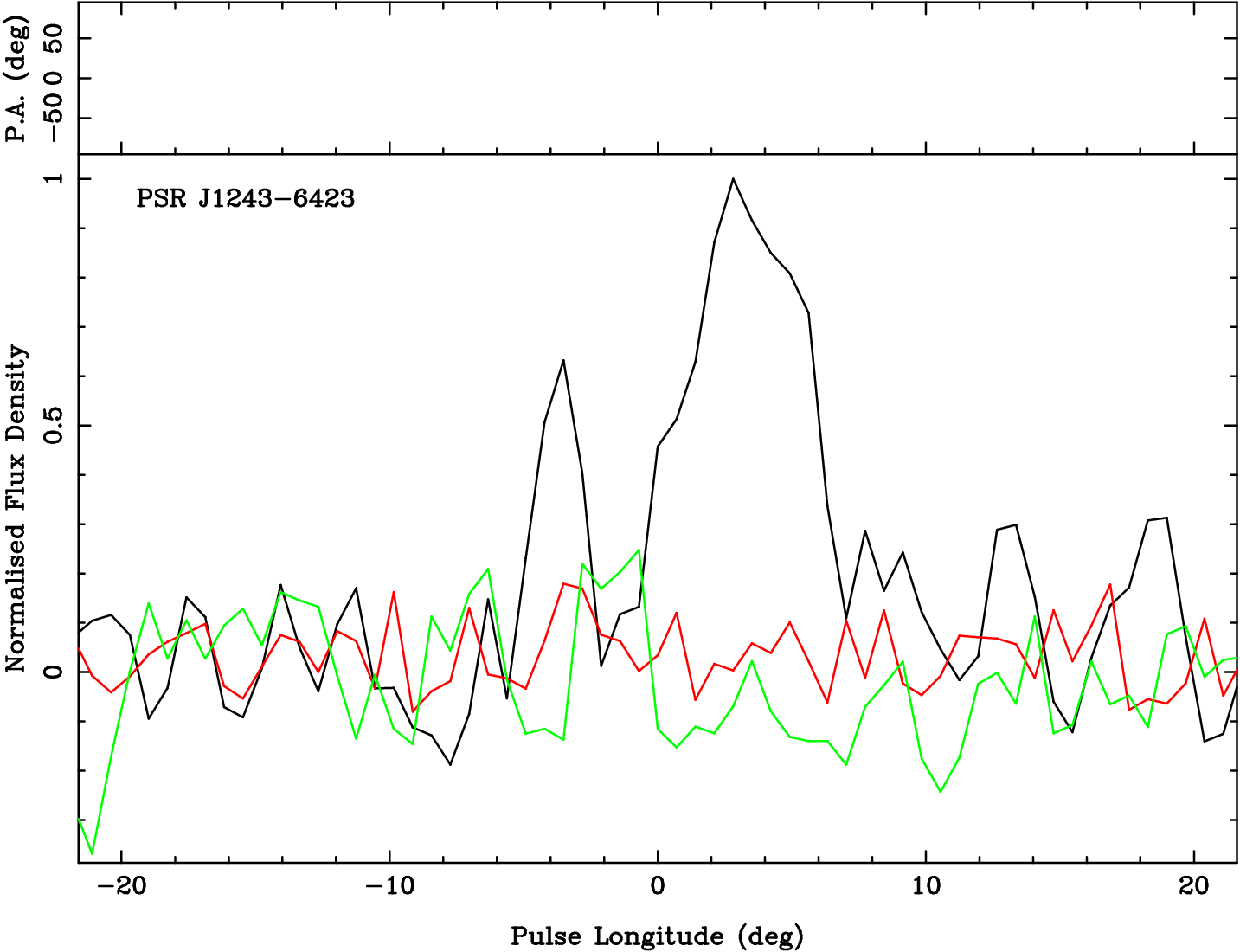}}&
\resizebox{0.4\hsize}{!}{\includegraphics[angle=-90]{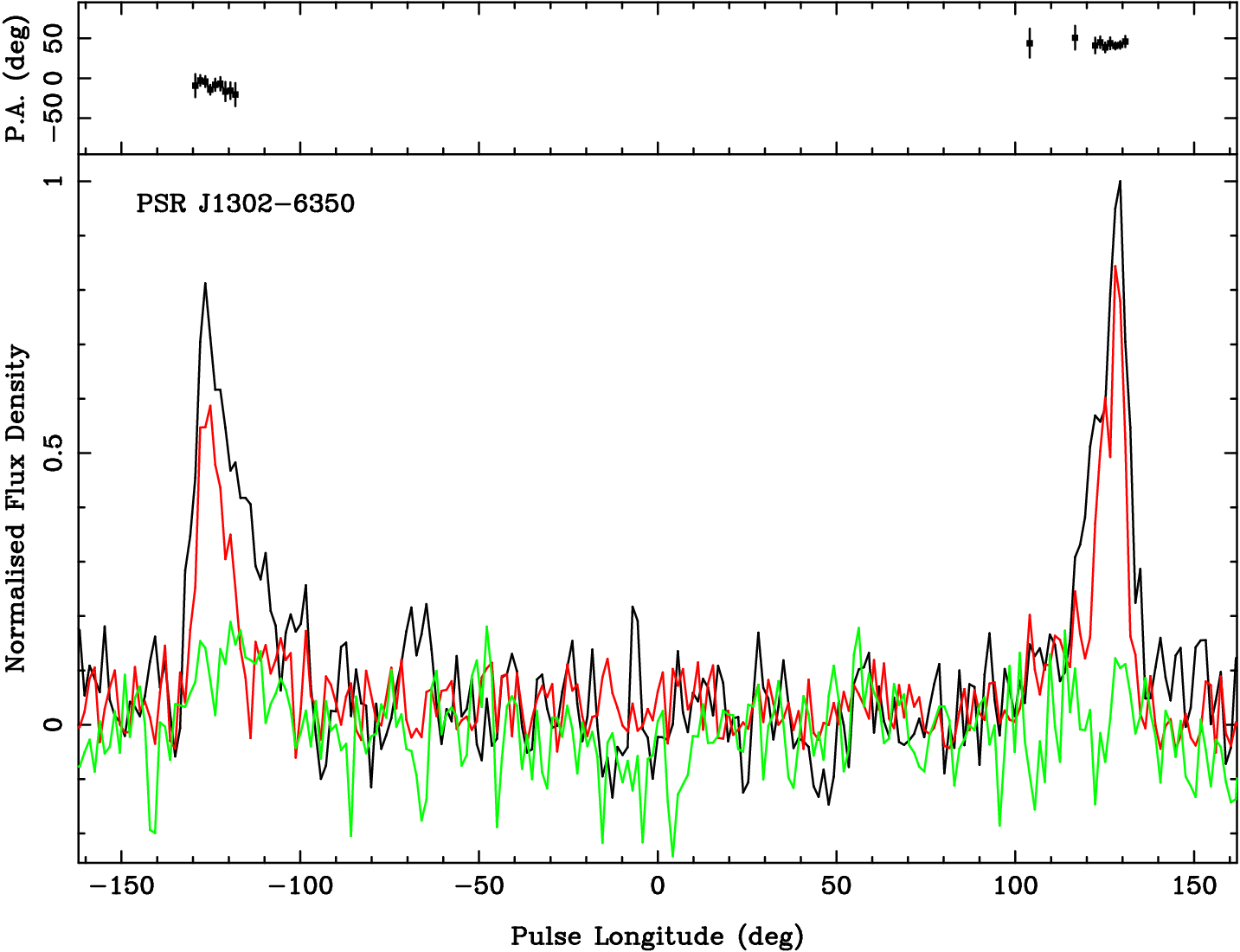}}\\
\resizebox{0.4\hsize}{!}{\includegraphics[angle=-90]{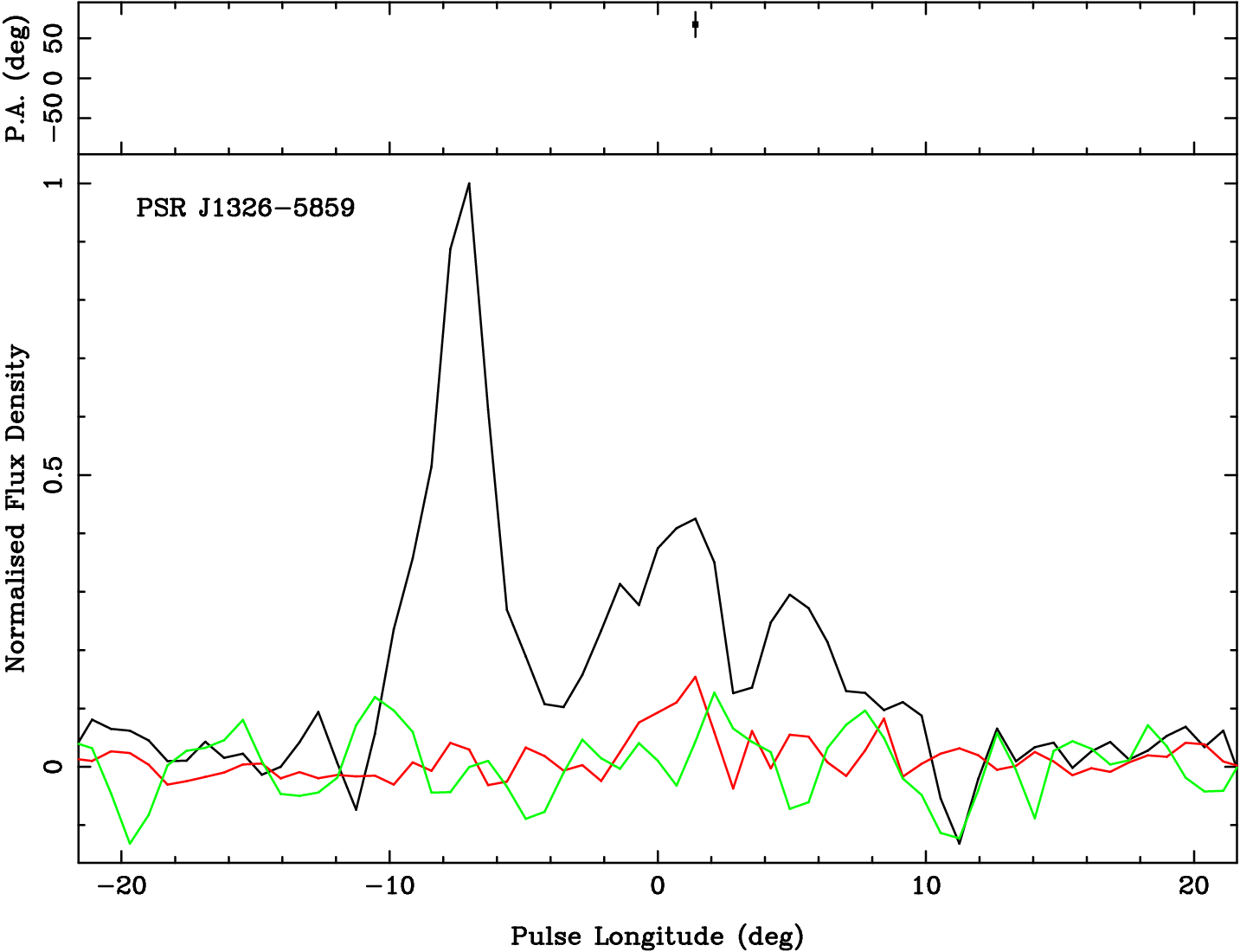}}&
\resizebox{0.4\hsize}{!}{\includegraphics[angle=-90]{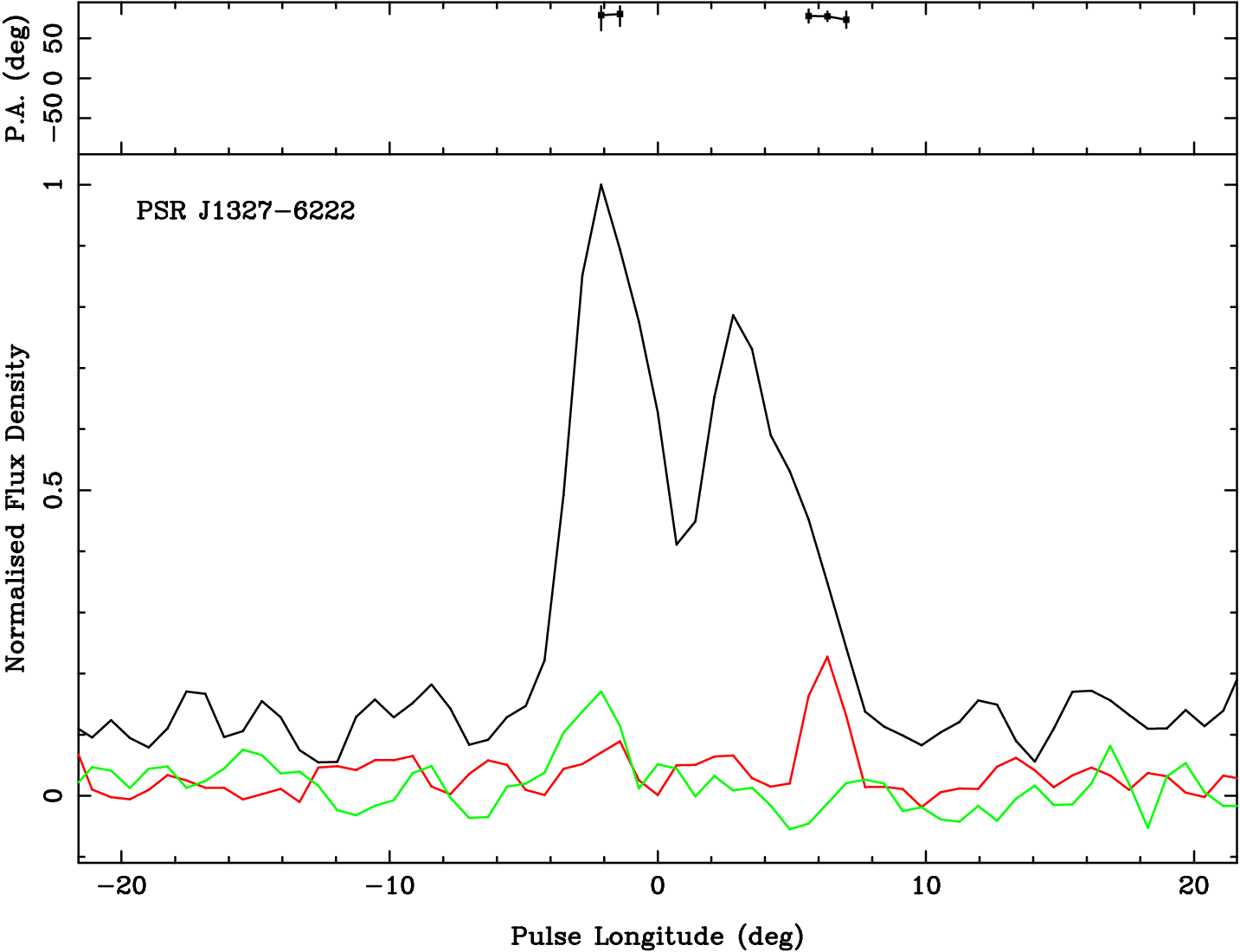}}\\
\end{tabular}
\end{center}
\caption{Polarisation profiles at 8.4~GHz for 8 pulsars as marked.
See Figure~1 for details.}
\end{figure*}

\noindent
{\bf PSR J0630$-$2834 (B0628$-$28):} Observations of this pulsar show
high linear polarization, modest negative circular polarization and a
smooth PA swing at frequencies between 0.41 and 1.6~GHz \citep{gl98}.
At 3.1~GHz the polarization fraction is significantly reduced and
there is no polarization at 4.85~GHz \citep{hkk98}.  Our 8.4~GHz
observations also show no linear polarization although there is
perhaps a hint of positive circular through the centre of the profile.
A general narrowing of the profile is observed from low to high
frequencies although the shape appears not to change. This profile
is likely to be a blend of conal components without a core.

\noindent
{\bf PSR J0659+1414 (B0656+14):} This pulsar has very high linear
polarization and significant negative circular polarization at
frequencies between 0.4 and 1.64~GHz \citep{gl98,wcl+99,wck+04}.  At
3.1~GHz the polarization remains high on the trailing half of the
pulse but is almost absent on the leading half. At 4.85~GHz the
polarization fraction has decreased significantly and an OPM jump is
visible on the leading part of the profile \cite{hoe99}.  Our 8.4~GHz
observations show that the profile has not evolved significantly and
there is a complete absence of both linear and circular polarization.
The pulse profile gets narrower with increasing frequency. This is one
of the few examples of a young pulsar which has become depolarized
at high frequencies. Its profile is perhaps a grazing conal component
like that of other, similar looking young pulsars \citep{jw06}.

\noindent
{\bf PSR J0738$-$4042 (B0736$-$40):} This pulsar has a highly complex
polarization pattern at 1.5 and 3.1~GHz (KJ06). There are two
orthogonal mode jumps close to the rising and falling edge of the
profile.  At 8.4~GHz the pulsar is much less polarized than at lower
frequencies.  The leading component retains some linear polarization
but the middle components have only a trace remaining. The circular
polarization remains low, as at lower frequencies. The orthogonal jump
near the rising edge of the profile appears to be a constant feature
at all frequencies.  The second orthogonal jump on the trailing edge
of the profile seen at low frequencies cannot be seen at 8.4~GHz due
to the low fractional polarization. It seems unlikely that this pulsar
contains a central core component as there is no great shape change
over a wide frequency range. We consider it likely that this is in the
class of symmetric conal structures with no core.

\noindent
{\bf PSR J0742$-$2822 (B0740$-$28):} At lower frequencies the pulsar
consists of as much as seven components \citep{kra94}, of which all
except the last are highly linearly polarized (KJ06). A previous
observation at 8.4~GHz was made by Morris et al. (1981)\nocite{mgs+81}
and, as in our observations, the leading
component now dominates the profile but the other components are still
clearly visible, particularly in circular polarization. The linear
polarization remains high throughout.  The OPM jump seen at the
trailing edge of the profile at lower frequencies cannot be discerned
here due to the low linear polarization.

The frequency evolution of the profile is at odds with what one might
expect. Figure~\ref{0742} shows the pulse profile at 4 different frequencies.
At 0.6~GHz the profile is rather asymmetric but it becomes more
symmetric at 1.4, 3.1 (KJ06) and 4.9~GHz \citep{hx97}.
The likely interpretation of the spectral index behaviour is that the profile
contains (at least) two outer conal rings with a central core components. Fits
to the rotating vector model also favour this solution \citep{jhv+05}.
At 8.4~GHz however, the trailing edge has dramatically declined, contrary
to expectations. The 10.6~GHz profile
of Xilouris et al. (1995)\nocite{xsg+95} shows similar features.
This implies there must be a strong spectral break in the trailing 
components at a frequency of $\sim$7~GHz.

\noindent
{\bf PSR J0835$-$4510 (B0835$-$45):} High time resolution observations
of this pulsar show that it consists of 3 main components at frequencies
above 1.4~GHz \citep{jhv+05}. Below this frequency scattering
dominates the profile shape. At 1.4~GHz the leading component
dominates the profile with the trailing component rather weak. At
3.1~GHz both the middle and trailing components are bigger with
respect to the initial component. At 8.4~GHz the trailing component is
now dominating the profile and the leading component has become
weak. Clearly then the initial component has a rather steep spectral
index in contrast to the other two components.  The linear
polarization remains very high between 1.4 and 8.4~GHz and the
circular polarization increases significantly with increasing
frequency in this young pulsar.
The PA swing appears to get steeper with increasing frequency.
The initial component is most likely to be the core, followed by trailing
edge cones with the leading cone not detected (see also
Johnston et al. 2001\nocite{jvkb01}).

\noindent
{\bf PSR J0837$-$4135 (B0835$-$41):} The frequency evolution of this
pulsar shows `classical' behaviour. At low frequencies, the central
component completely dominates the profile and the linear polarization
is relatively high. As one goes to higher frequencies, the conal
outriders gradually become more prominent and the linear polarization
declines. The circular polarization remains virtually constant over a
large frequency range.  The PA in the outer components is the same as
that seen at lower frequencies (KJ06) but the complex PA swing across
the centre of the pulse at lower frequencies cannot be traced at this
frequency. Clearly this is a symmetric profile with a central core.

\noindent
{\bf PSR J0908$-$4913 (B0906$-$49):} High time resolution observations
of this pulsar show that both the main and interpulses consist of two
components. The two components in the interpulse maintain their
intensity ratio between 0.66 and 8.4~GHz. Both components are
completely linearly polarized at all frequencies.  In the main pulse,
the two components are blended together; both are highly linearly
polarized and the second component also has strong circular
polarization which increases with increasing frequency.  The ratio of
the two components decreases with increasing frequency with the
trailing component dominating the profile. This is a young pulsar
and both the main and interpulses are symmetric doubles with no core
as seen in other pulsars of this type \citep{jw06}.

\noindent
{\bf PSR J0922+0638 (B0919+06):} At frequencies between 0.4 and
1.6~GHz, the profile of the pulsar consists of a weak leading
component and a strong trailing component which is highly linearly
polarized \citep{gl98}.  The total intensity profile is similar at the
low frequency of 47~MHz \citep{pw92}.  The leading component gets
weaker with increasing frequency; it is barely detectable at 3.1~GHz and
not present at 4.8~GHz \citep{hx97}.  The linear polarization is high
at all frequencies.  In our 8.4~GHz observations the linear
polarization has decreased and appears to have shifted towards the
early part of the profile unlike at lower frequencies.
This asymmetric pulse profile is likely a trailing edge cone.

\noindent
{\bf PSR J0953+0755 (B0950+08):} This well known pulsar has been
extensively studied over a wide frequency range (see the discussion in
Everett \& Weisberg 2001\nocite{ew01}), including frequencies as low
as 25~MHz \citep{pw92}. A low fraction of linear polarization is
present up to 5~GHz. Unfortunately the pulsar is rather weak at 8.4~GHz and
has lost virtually all its polarization.  However, the overall pulse
shape is similar to that at lower frequencies.  The interpulse is not
visible in our data, likely because of low signal to noise. Debate continues
as to whether the profile is consistent with a wide double profile or
whether the interpulse emission originates from a different pole to the
main pulse (see Everett \& Weisberg 2001\nocite{ew01}). We favour the
main pulse being a trailing edge cone.

\noindent
{\bf PSR J1048$-$5832 (B1046$-$58):} This is a young pulsar which shows
two distinct components at 1.4~GHz with the leading component being
highly linearly polarized and with moderate circular polarization.
At 3.1~GHz the polarization remains high but the trailing component
has dropped in amplitude with respect to the leading component (KJM05).
At 8.4 GHz, the trailing component has virtually disappeared but
now a highly polarized leading component can more clearly be 
discriminated than at lower frequencies. This is therefore a leading
edge cone, perhaps with a core at lower frequencies.
\begin{figure*}
\begin{center}
\begin{tabular}{cc}
\resizebox{0.4\hsize}{!}{\includegraphics[angle=-90]{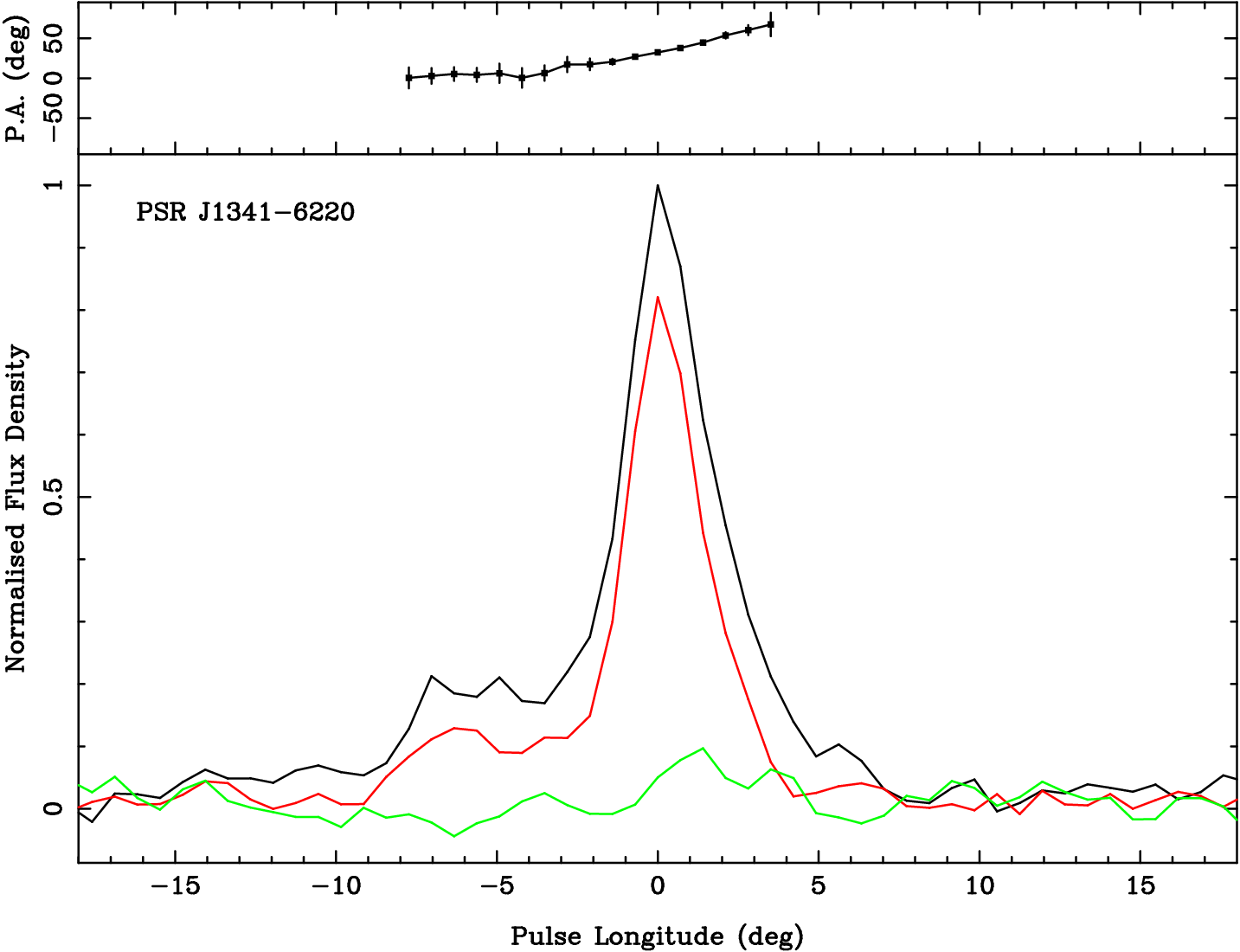}}&
\resizebox{0.4\hsize}{!}{\includegraphics[angle=-90]{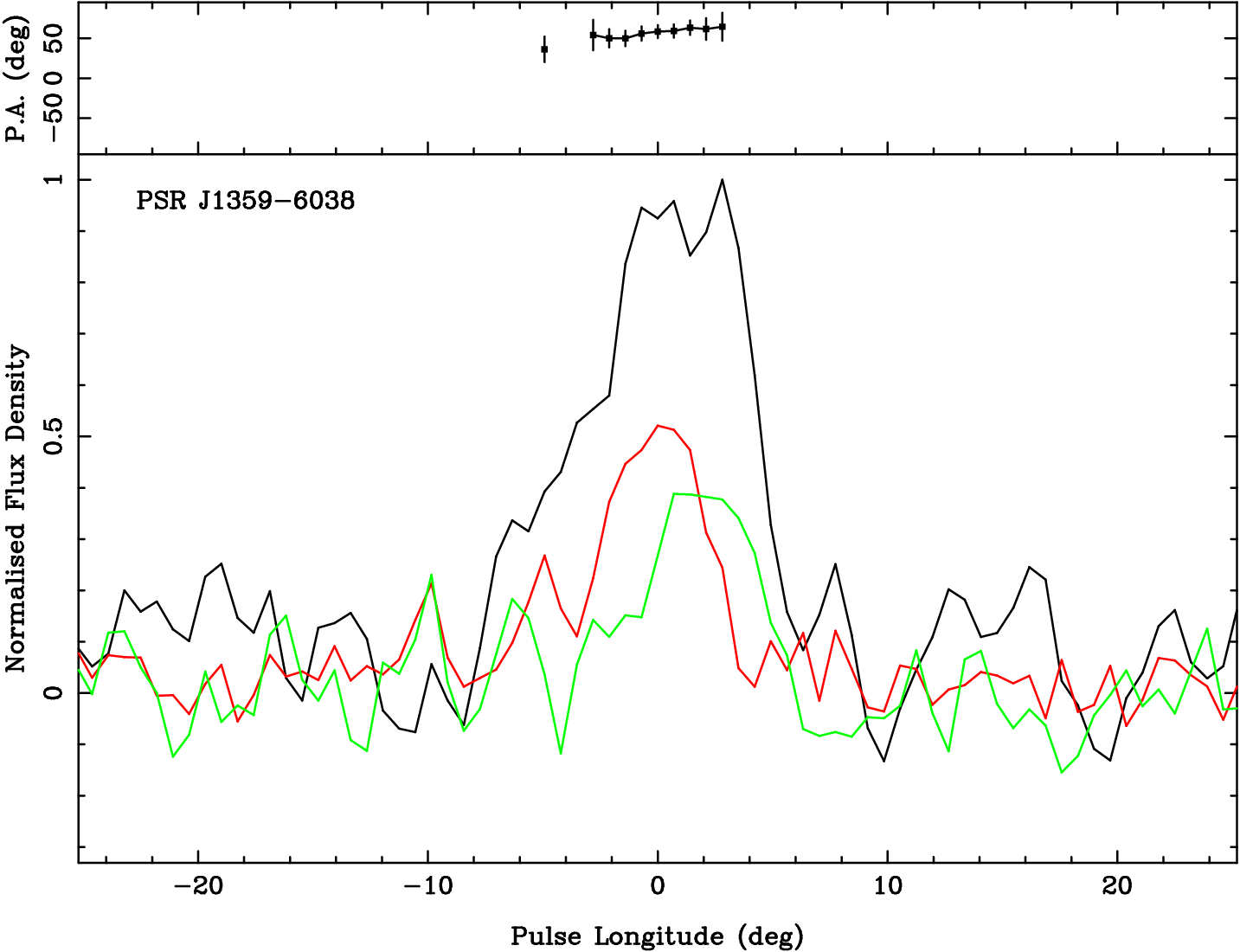}}\\
\resizebox{0.4\hsize}{!}{\includegraphics[angle=-90]{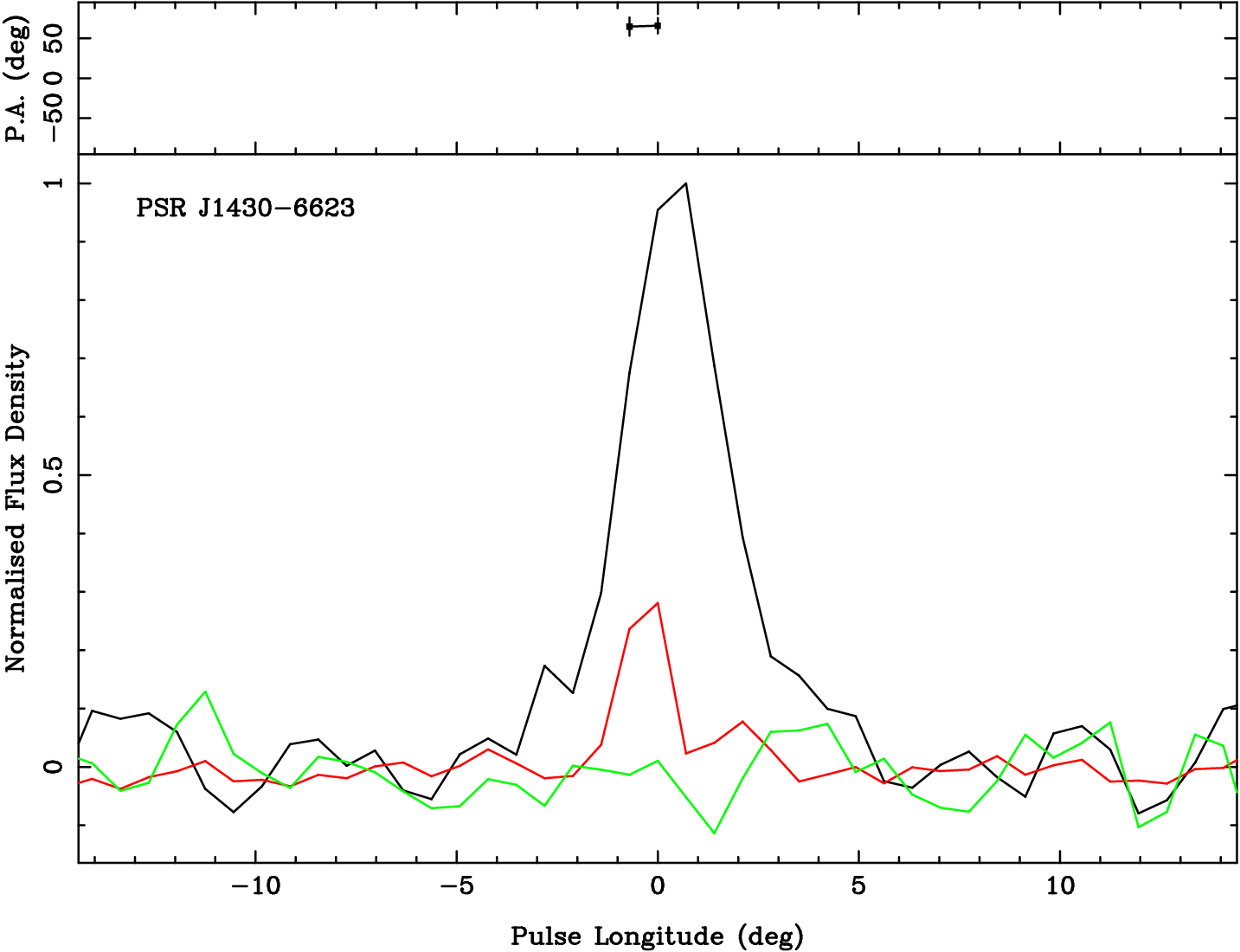}}&
\resizebox{0.4\hsize}{!}{\includegraphics[angle=-90]{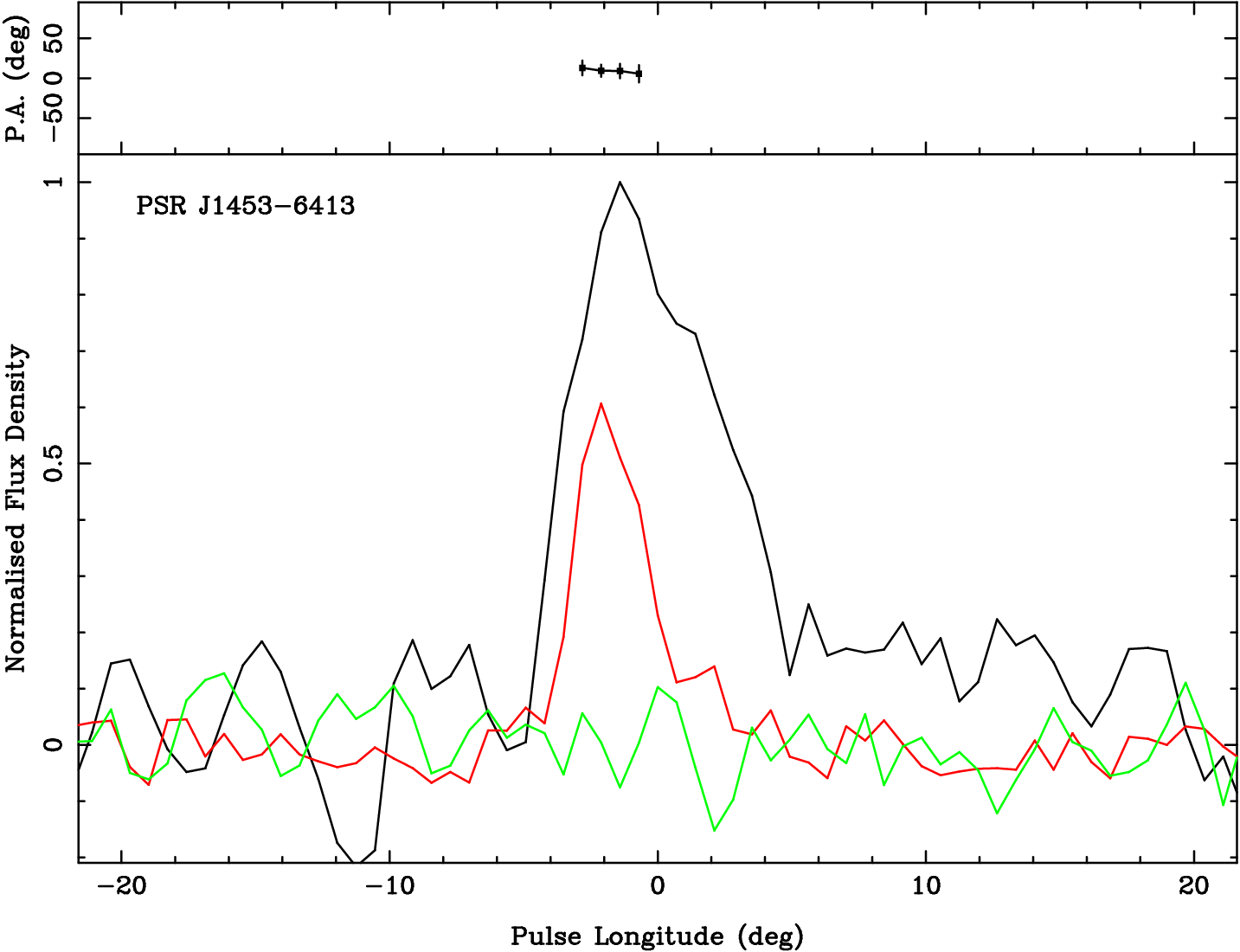}}\\
\resizebox{0.4\hsize}{!}{\includegraphics[angle=-90]{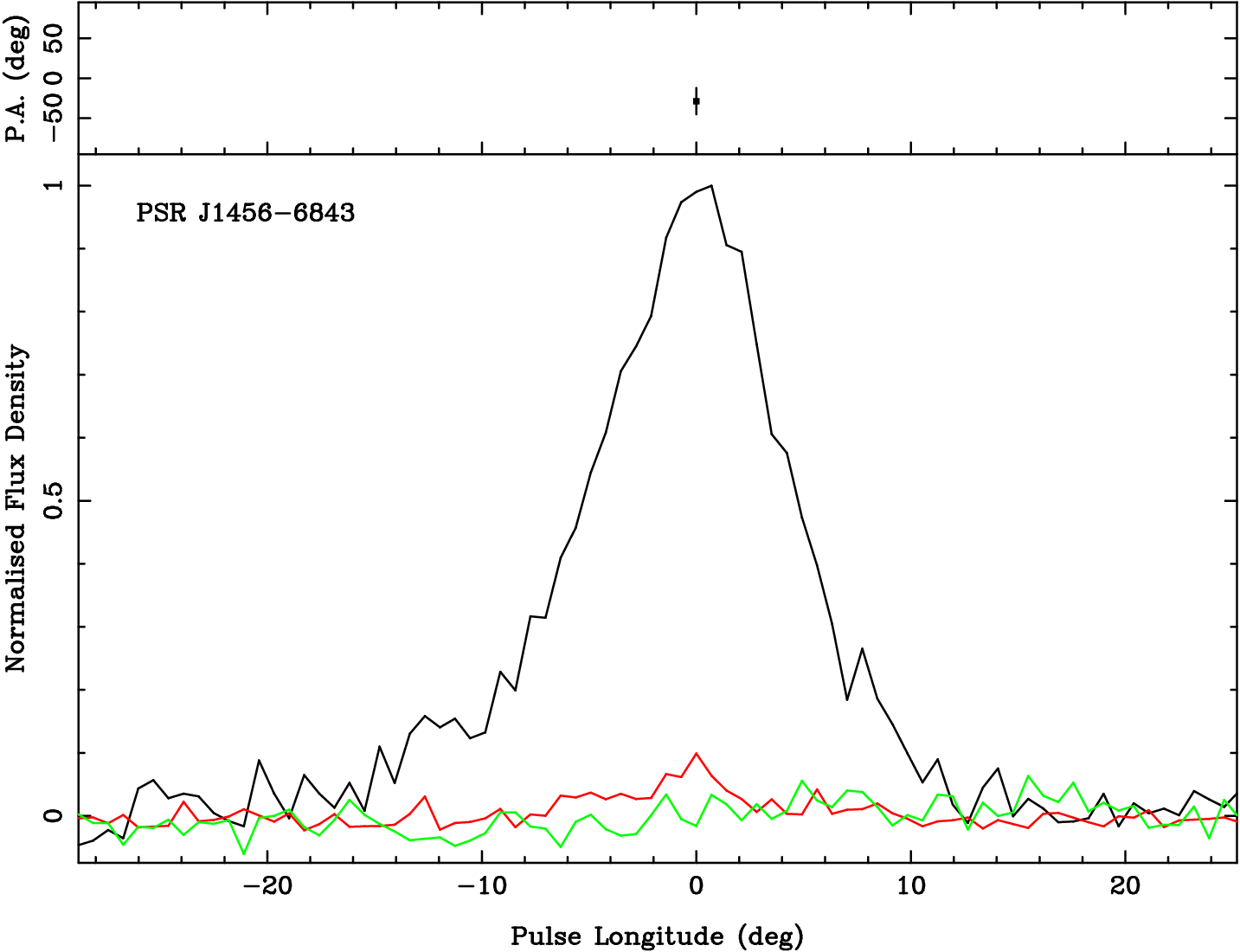}}&
\resizebox{0.4\hsize}{!}{\includegraphics[angle=-90]{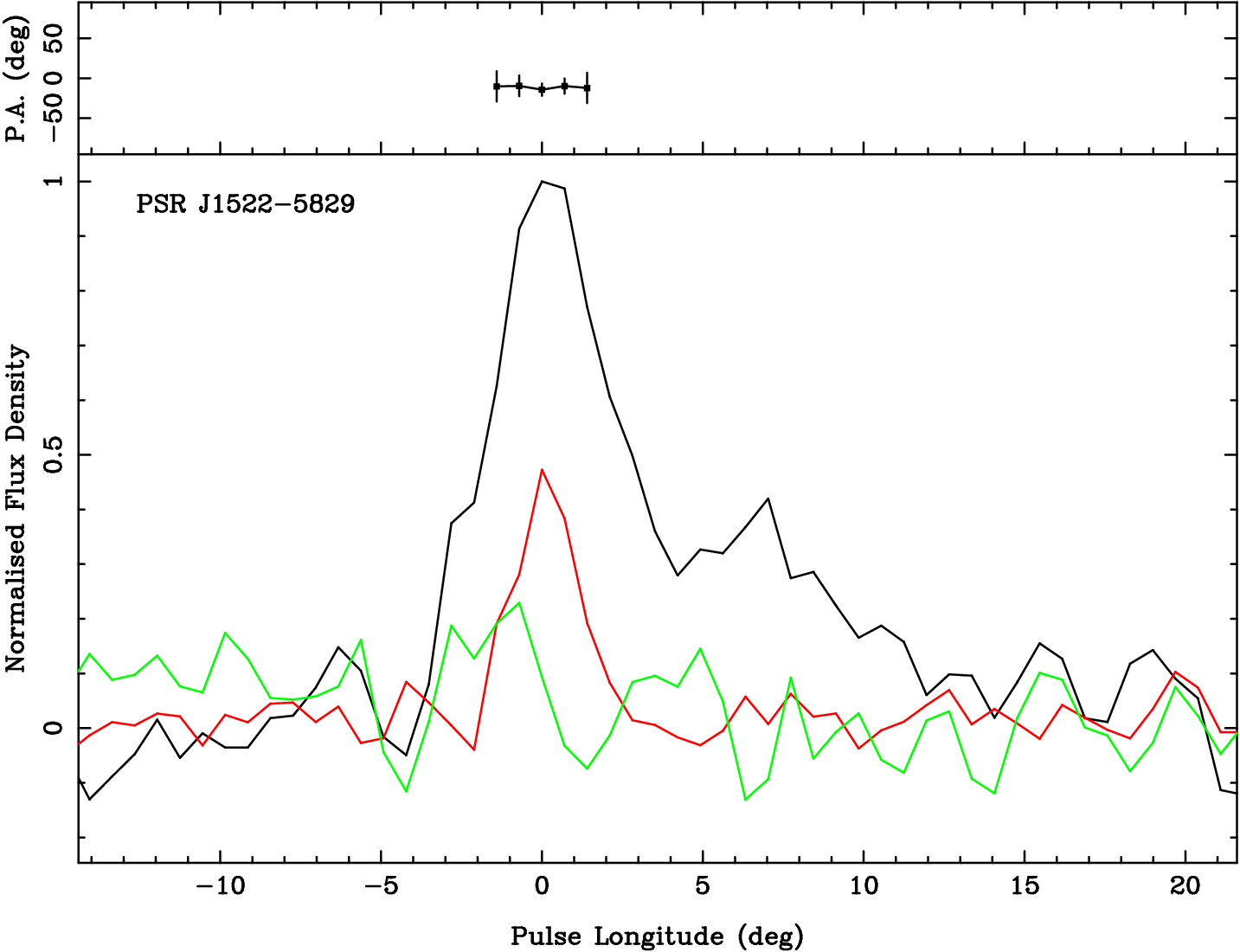}}\\
\resizebox{0.4\hsize}{!}{\includegraphics[angle=-90]{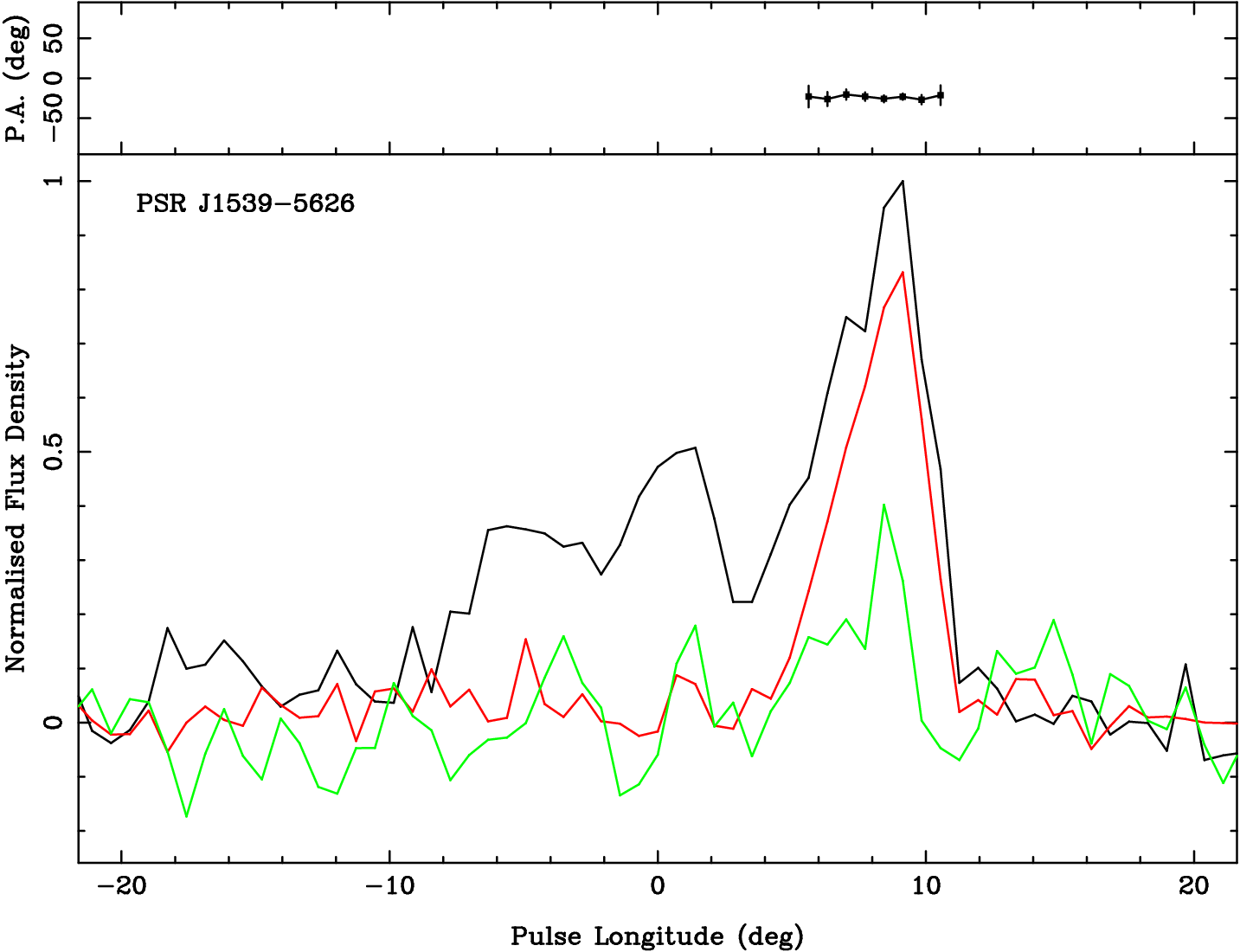}}&
\resizebox{0.4\hsize}{!}{\includegraphics[angle=-90]{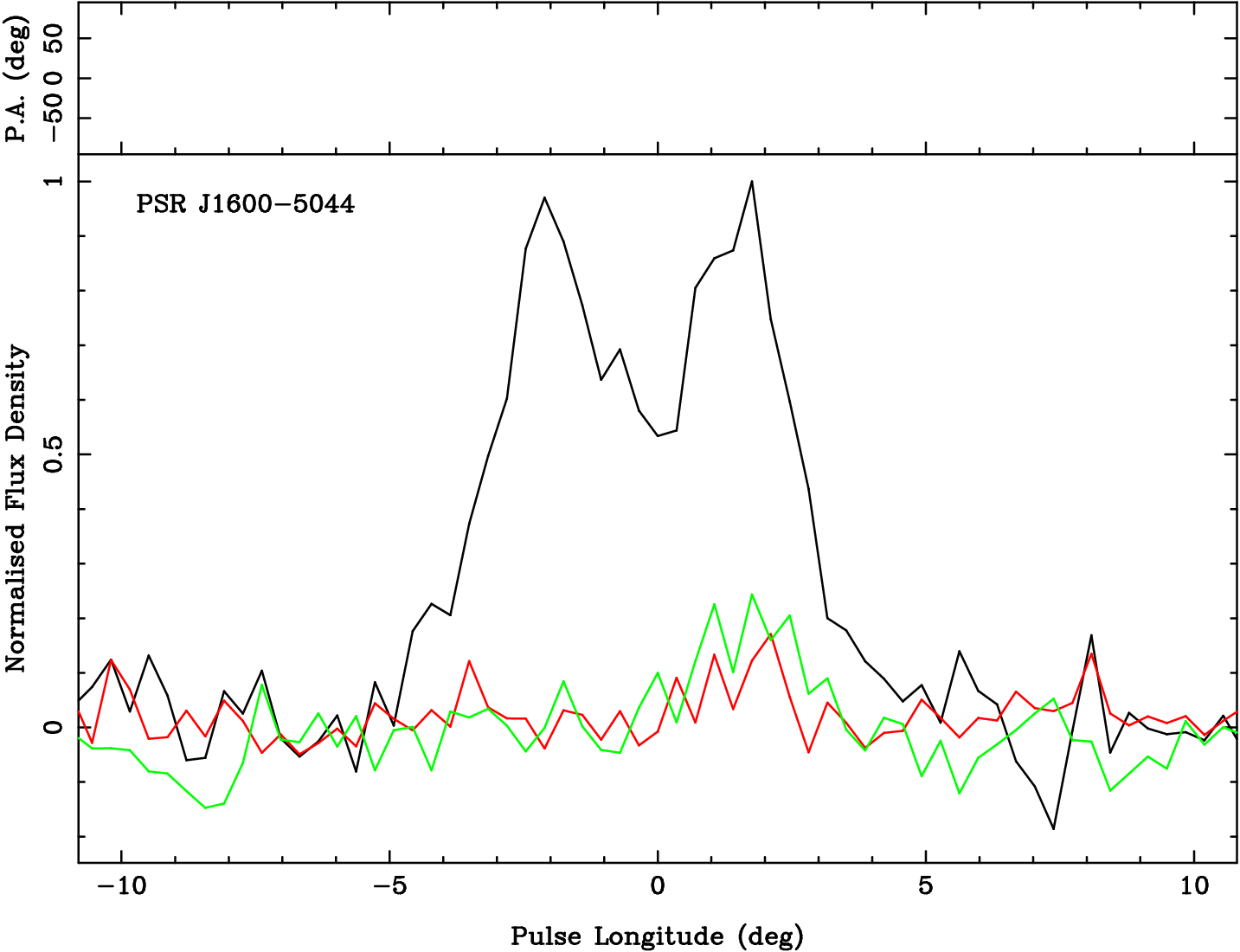}}\\
\end{tabular}
\end{center}
\caption{Polarisation profiles at 8.4~GHz for 8 pulsars as marked.
See Figure~1 for details.}
\end{figure*}
\begin{figure*}
\begin{center}
\begin{tabular}{cc}
\resizebox{0.4\hsize}{!}{\includegraphics[angle=-90]{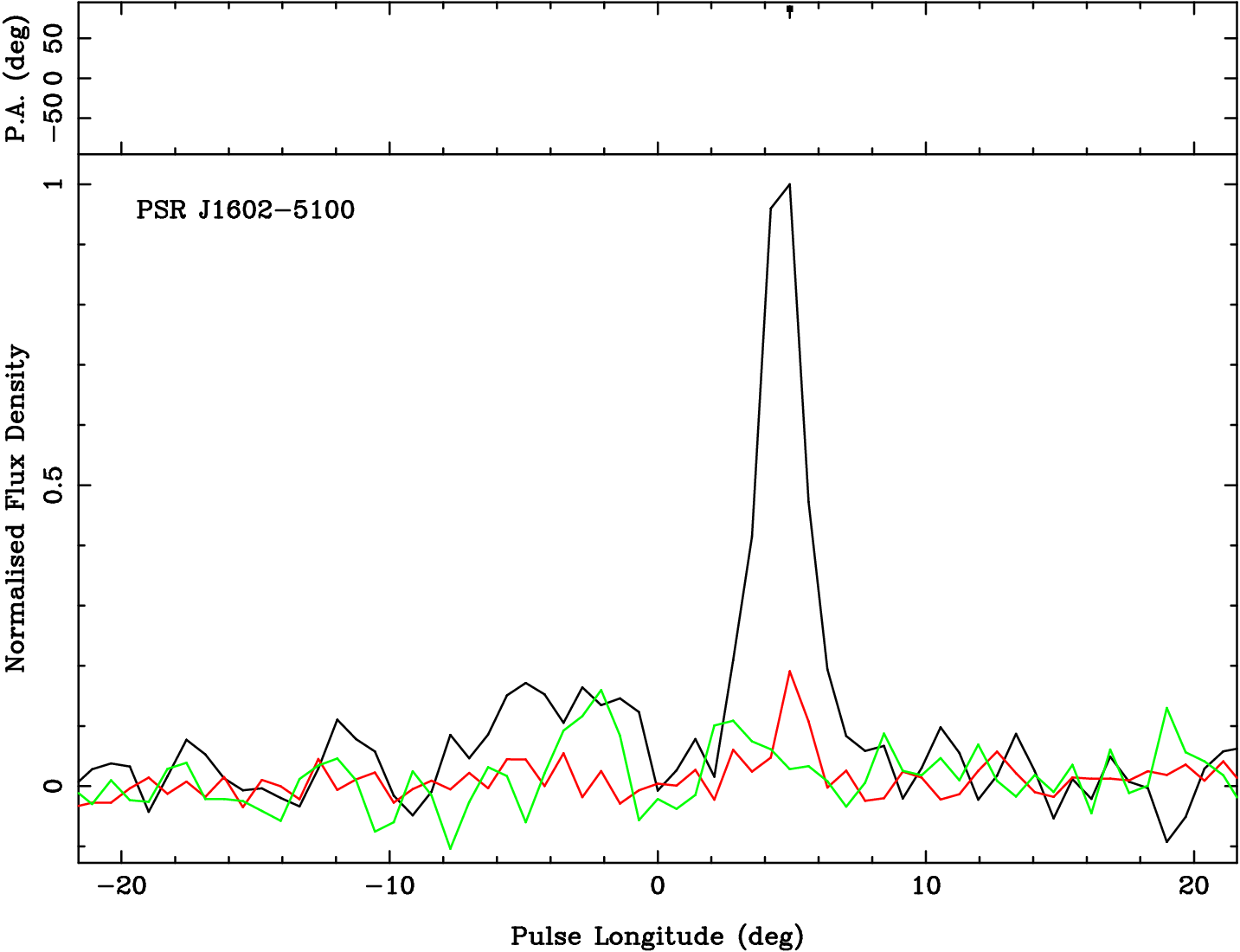}}&
\resizebox{0.4\hsize}{!}{\includegraphics[angle=-90]{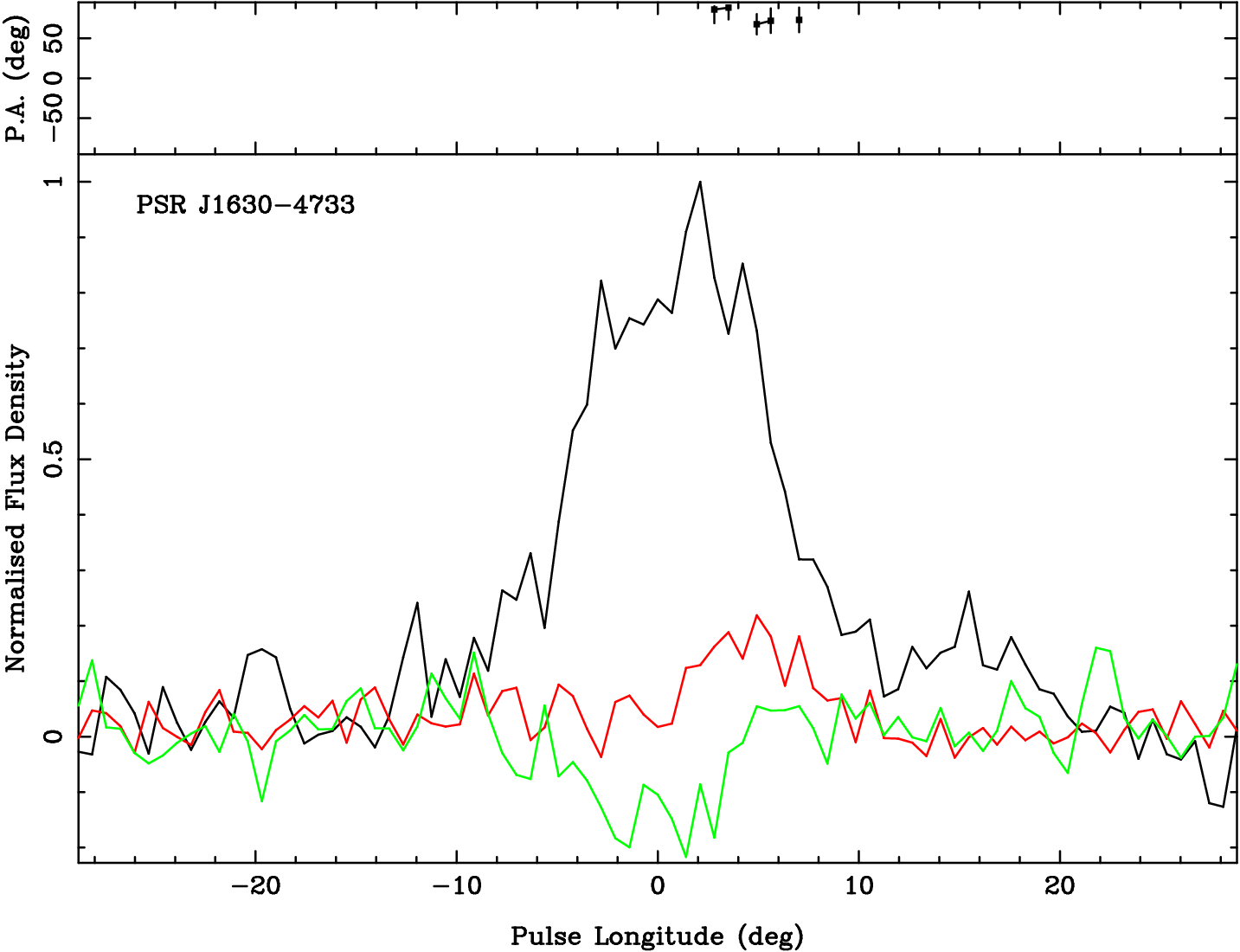}}\\
\resizebox{0.4\hsize}{!}{\includegraphics[angle=-90]{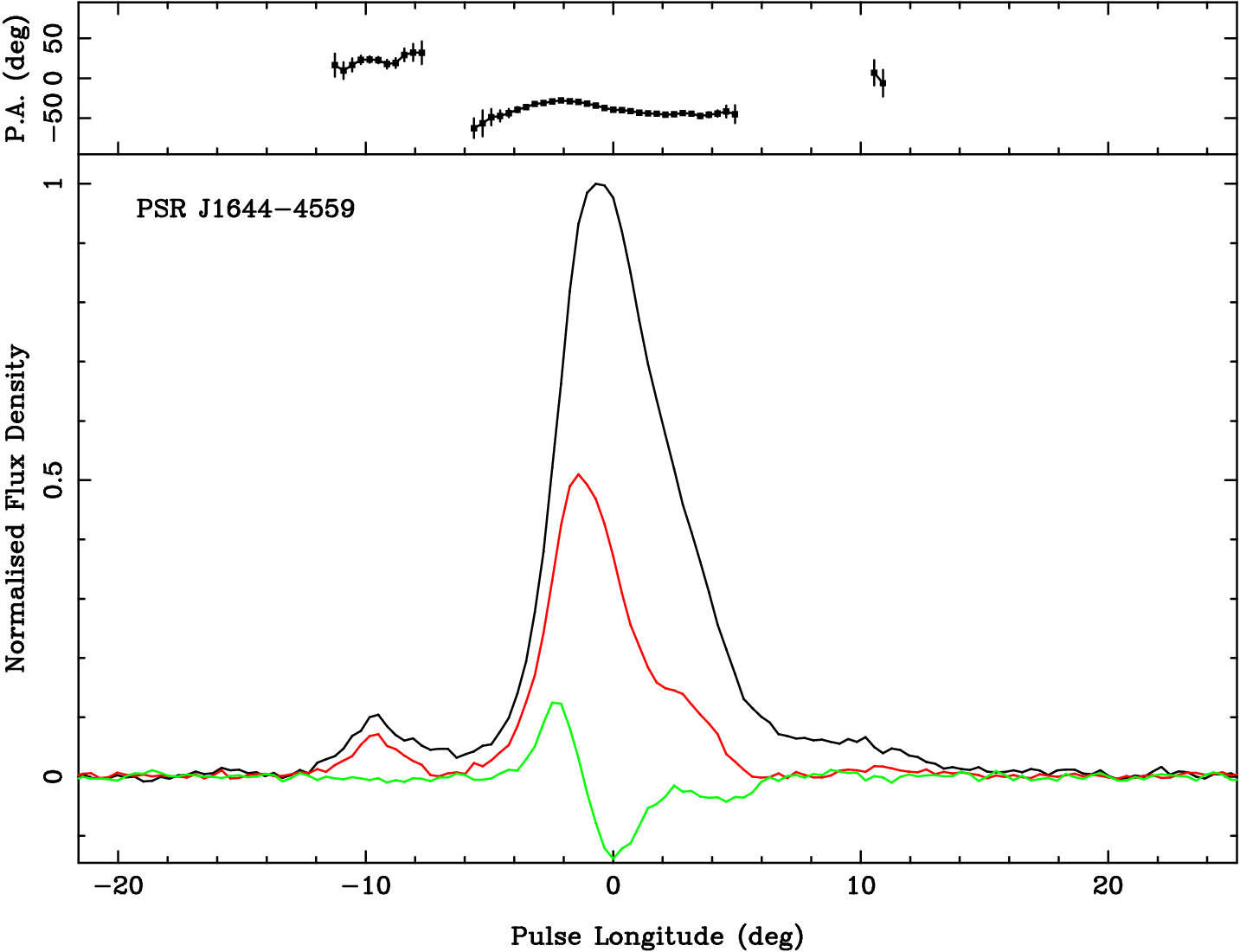}}&
\resizebox{0.4\hsize}{!}{\includegraphics[angle=-90]{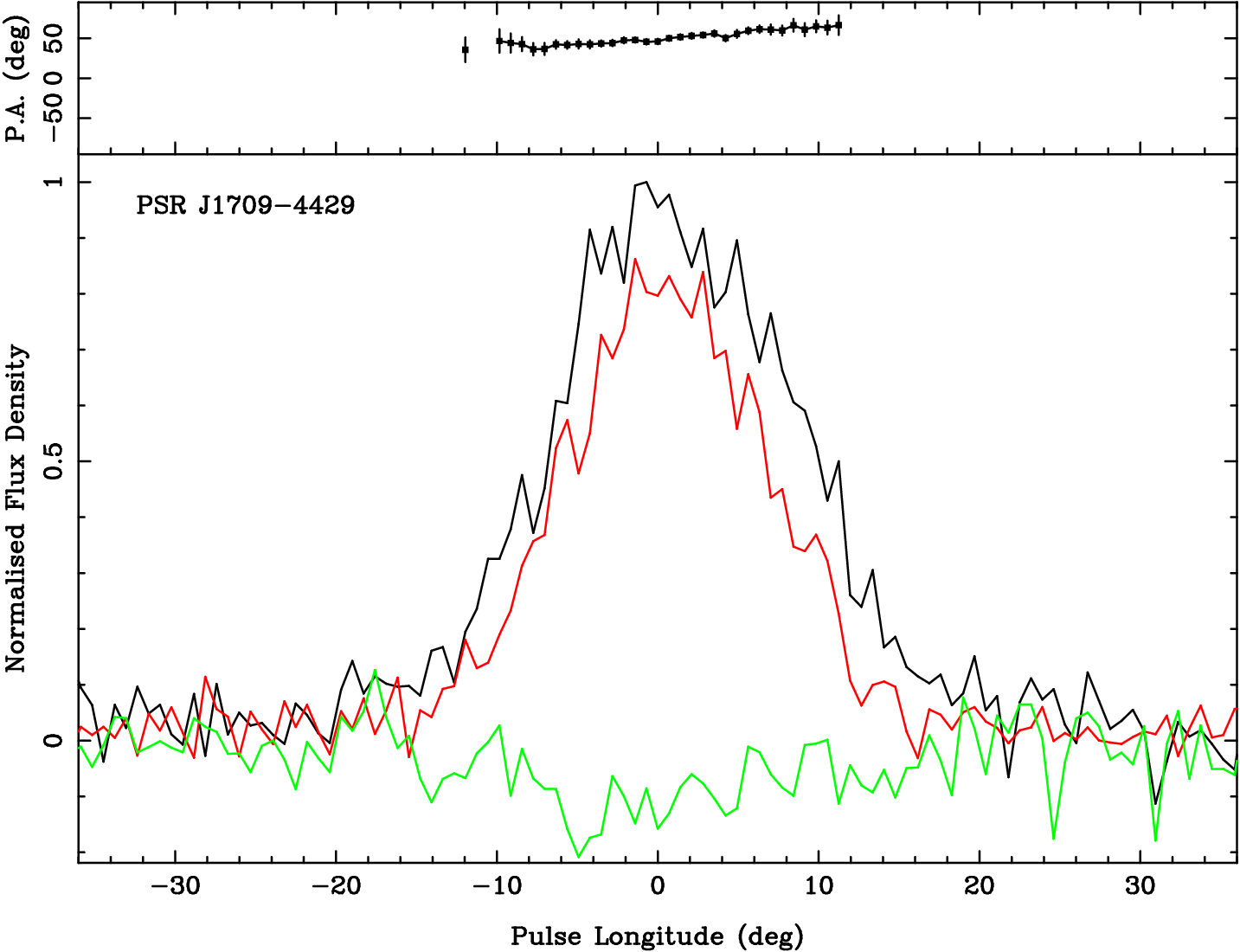}}\\
\resizebox{0.4\hsize}{!}{\includegraphics[angle=-90]{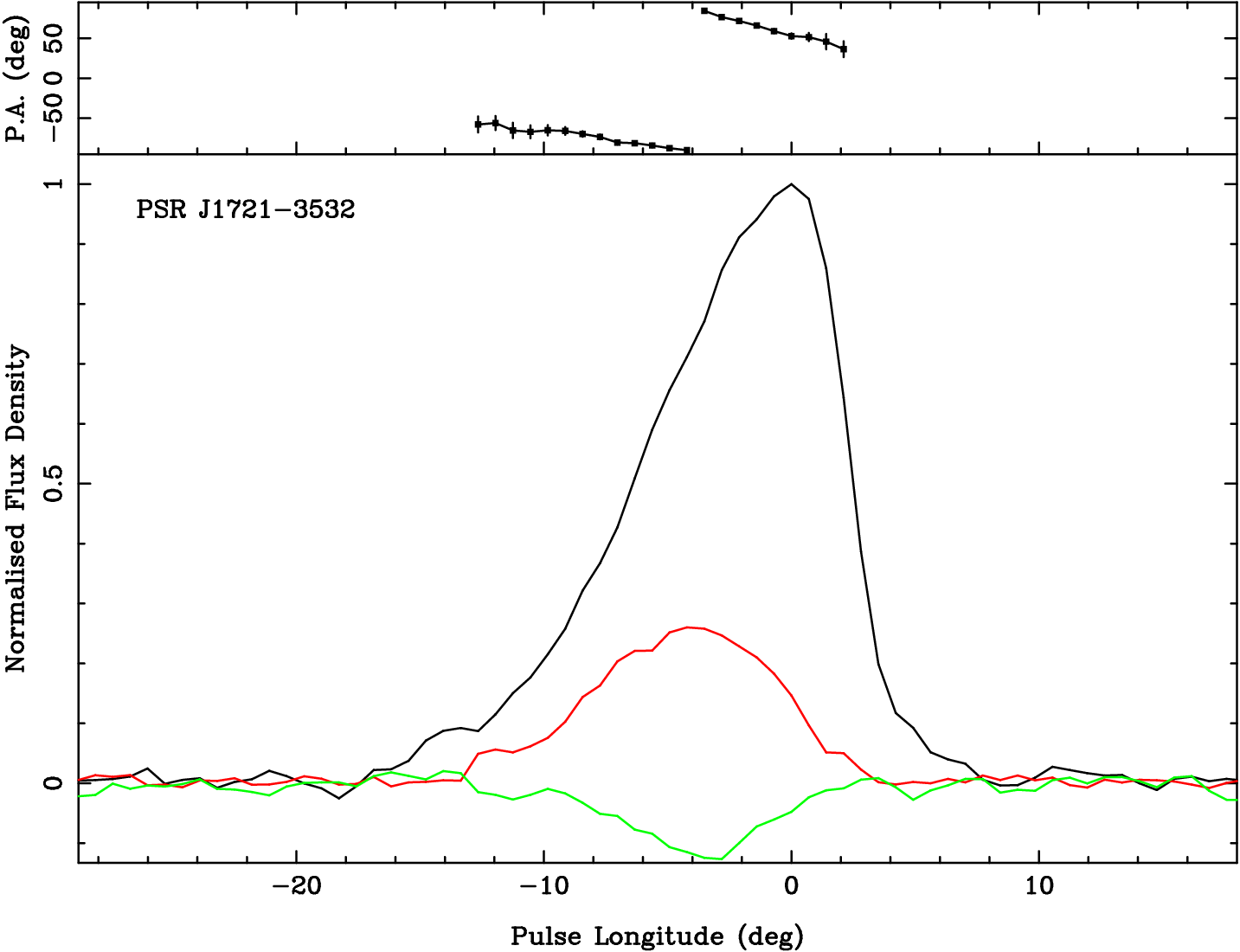}}&
\resizebox{0.4\hsize}{!}{\includegraphics[angle=-90]{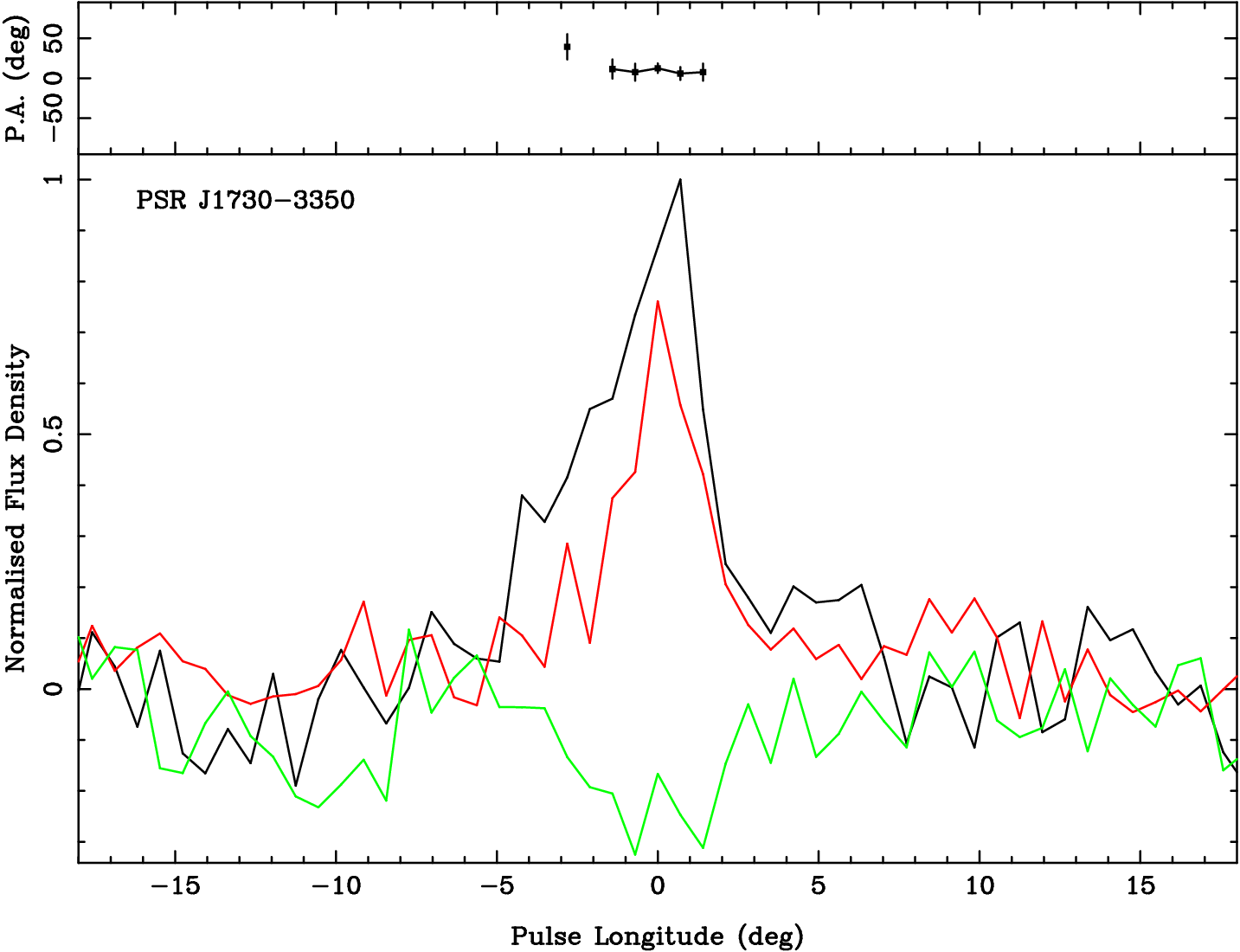}}\\
\resizebox{0.4\hsize}{!}{\includegraphics[angle=-90]{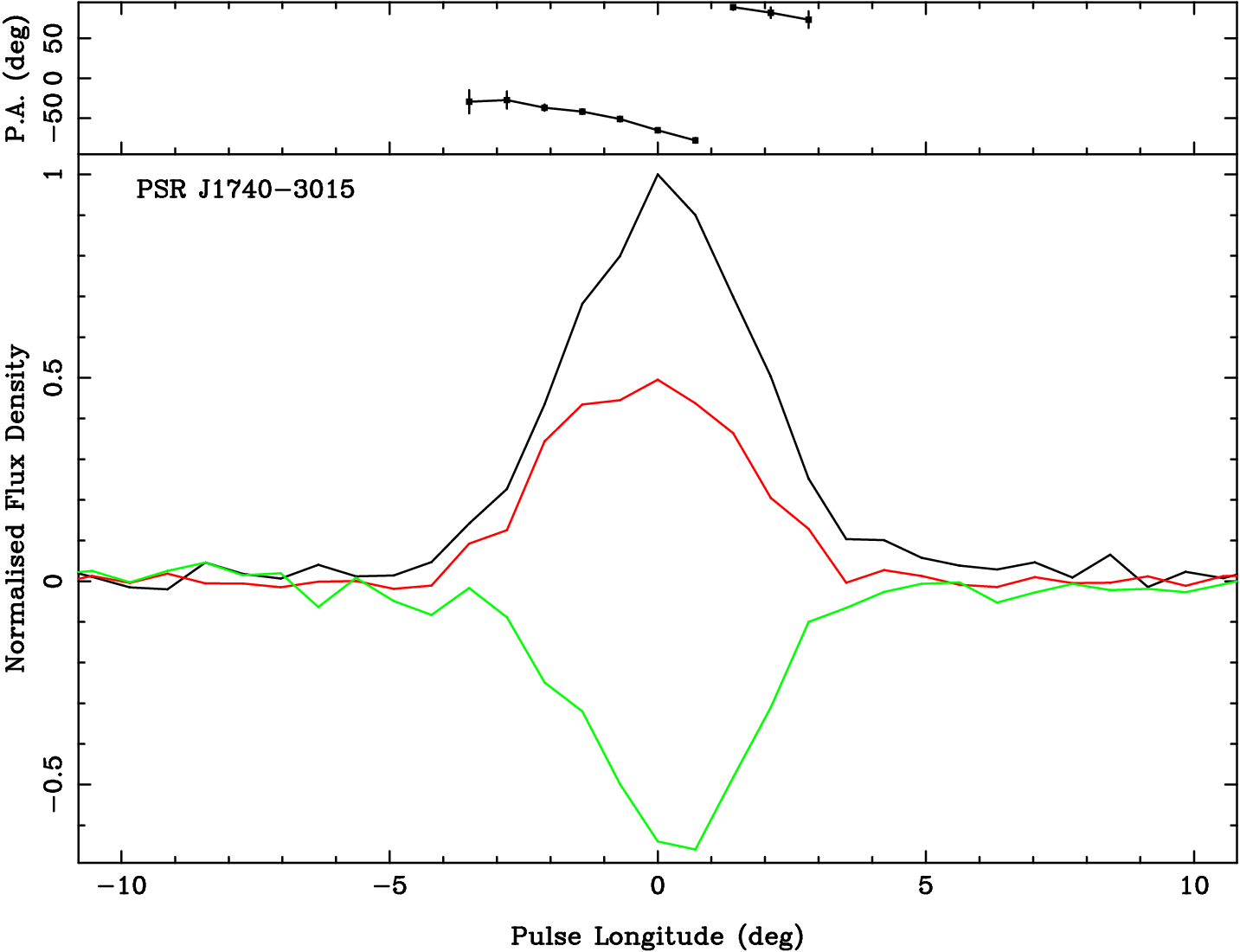}}&
\resizebox{0.4\hsize}{!}{\includegraphics[angle=-90]{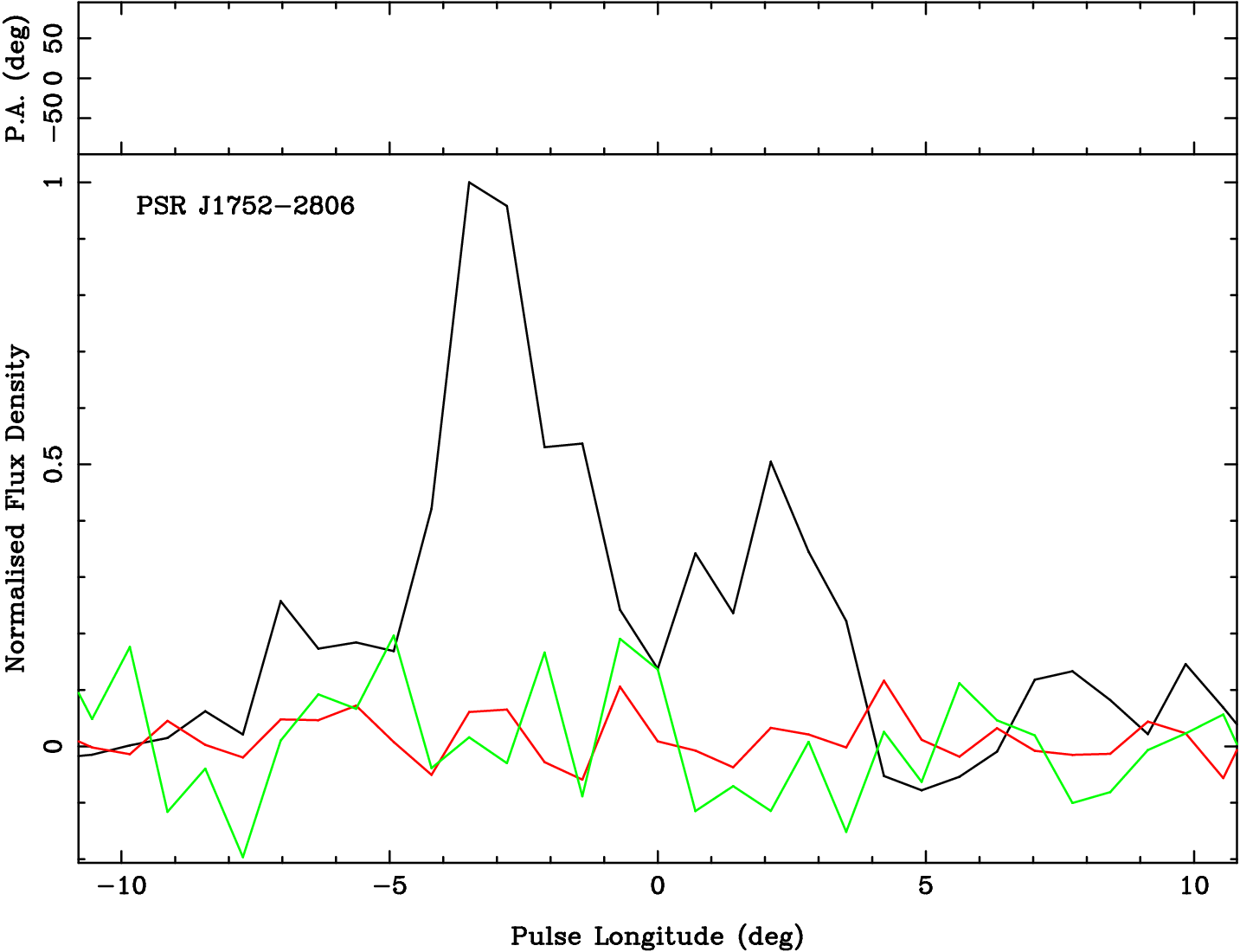}}\\
\end{tabular}
\end{center}
\caption{Polarisation profiles at 8.4~GHz for 8 pulsars as marked.
See Figure~1 for details.}
\end{figure*}

\noindent
{\bf PSR J1056$-$6258 (B1054$-$62):} The linear polarization of this
pulsar is relatively high at 1.4~GHz but is already in decline by
3.1~GHz (KJ06). In the current 8.4~GHz observations the polarization
is virtually absent.  Although the PA swings at 1.4 and 3.1~GHz are
different, the lack of polarization at 8.4~GHz precludes any further
comment on the frequency behaviour.  The total intensity profile looks
similar to that at lower frequencies.
It seems likely that this is a (leading-edge) partial cone with 
the magnetic pole crossing at later longitudes.

\noindent
{\bf PSR J1136+1551 (B1133+16):} This well known pulsar has been
extensively studied over a large frequency range and shows a classic
double pulse profile. At very low frequencies, the components have
almost equal strength but the leading component has a significantly
flatter spectral index than the trailing component and dominates at
high frequencies. The linear polarization is reasonably high at
0.4~GHz but has dropped considerably at 1.4 and 4.8~GHz due to
competition between orthogonal modes as revealed in the single pulse
study of Karastergiou et al. (2002)\nocite{kkj+02}.  In our 8.4~GHz
observations, and also at 10.5~GHz \citep{hx97} a trace of polarization
remains in the leading component.

\noindent
{\bf PSR J1243$-$6423 (B1240$-$64):} At low frequencies this profile
is a simple Gaussian. The circular polarization swings through the
centre of the pulse and the PA swing is also steep (van Ommen et
al. 1997\nocite{vdhm97}, KJ06). This indicates a core component.
At 3.1~GHz conal
outriders are starting to appear on each side of the central component
(KJ06), with the leading conal component being stronger and narrower
than the trailing component.  At 8.4~GHz the signal to noise is low
and no polarization can be seen. However, the leading conal component
has increased significantly in strength compared to the central
component. The trailing component however does not appear to have the
same frequency evolution.

\noindent
{\bf PSR J1302$-$6350 (B1259$-$63):} This is the well known pulsar
with a Be star companion whose total intensity profiles at a range of
frequencies were most recently shown in Wang, Johnston \& Manchester
(2004)\nocite{wjm04}.  At 8.4~GHz the individual components are somewhat 
narrower than at lower frequencies. The steep rising edge and shallower decline
are also consistent with lower frequencies. The degree of linear
polarization remains high at 8.4~GHz, although there is a significant
decrease in polarization in the inner parts of both components.
The morphology and polarization of this profile are best described
by a wide double \citep{mj95} similar to other young pulsars.

\noindent
{\bf PSR J1326$-$5859 (B1323$-$56):} The profile of this pulsar
undergoes very strong frequency evolution. At 1.4~GHz the profile
consists mainly of a strong central component with rather weak conal
outriders. At 3.1~GHz the leading outrider has become significantly
more prominent (KJ06). In our 8.4~GHz profile, the leading component
now dominates the profile and the central component is significantly
reduced.  The ratio of the amplitudes of the leading and trailing
component changes little between 3.1 and 8.4~GHz.  The polarization is
complex at low frequencies with a swing of circular polarization
across the central component and a complicated PA swing (KJ06). At
8.4~GHz the polarization is almost completely absent. This is clearly
a symmetrical profile with a central core.

\noindent
{\bf PSR J1327$-$6222 (B1323$-$62):} The profile at 8.4~GHz continues
the frequency evolution seen between 1.4 and 3.1~GHz (KJ06).  Whereas
the trailing component is dominant at low frequencies, this has been
reversed at 8.4~GHz. There is little polarization at this frequency
with which to compare to the lower frequency observations. There is little
or no core component present, this profile is a symmetrical cone.

\noindent
{\bf PSR J1341$-$6220 (B1338$-$62):} This is a young pulsar which
shows characteristic traits of relatively flat spectral index, a high
degree of linear polarization and a double pulse profile with the
trailing component dominating \citep{jw06}.  The
profile at 8.4~GHz is similar to that at 3.1~GHz with continued high
linear polarization; at lower frequencies the profile is heavily
scatter-broadened \citep{jw06}.

\noindent
{\bf PSR J1359$-$6038 (B1356$-$60):} At 1.4~GHz the profile appears to
consist of a single component which is highly linearly polarized. At
3.1~GHz a trailing component emerges and the polarization remains high
(KJ06).  At 8.4~GHz the linear polarization has declined somewhat and
the circular polarization has increased. A comparison between the PA
swing at 3.1 and 8.4~GHz shows almost perfect agreement in the regions
where there is overlap. This is most likely a leading edge conal
component.

\noindent
{\bf PSR J1430$-$6623 (B1427$-$66):} There is considerable frequency
evolution of the profile of this pulsar between 0.4 and 8.4~GHz.  At
frequencies of 1.4~GHz and below there is a prominent, wide leading
component and a narrow dominant trailing component (Hamilton et
al. 1977\nocite{hmak77}; Johnston et al. 2005).
At 3.1~GHz the leading component is
now substantially weaker. Both 1.4 and 3.1~GHz show identical and
complex polarization structure. There are at least three distinct
linear polarization features and the circular polarization changes
sign under the narrow trailing component. The PA swing is highly
complex and does not conform to the rotating vector model.  In these
8.4~GHz observations the leading component is now absent and only the
trailing component is detected. It retains its linear polarization and
the swing of circular polarization may also still be present.
This is therefore a trailing edge cone, with the core component absent
at high frequencies.

\noindent
{\bf PSR J1453$-$6413 (B1449$-$64):} There is substantial evolution of
the profile of this pulsar between 0.6 and 8.4~GHz.  At 0.6~GHz the
profile consists of a simple Gaussian component with moderate
polarization and a steep swing of position angle \citep{mhma78}.
At 1.4~GHz, a small leading component appears which is
highly polarized and the dominant trailing component has moderate
polarization orthogonal to the leading component. There is also a
highly extended tail which has a further orthogonal jump (Johnston et
al. 2005). At 3.1~GHz the leading component is now about as dominant
as the trailing component, there are no orthogonal jumps
and the extended tail is still present. In these 8.4~GHz observations,
the leading component entirely dominates the profile and remains
highly polarized. There is still a hint of an extended tail to the
profile. This is best explained by a leading edge cone plus weak core
component.

\noindent
{\bf PSR J1456$-$6843 (B1451$-$68):} The properties of this pulsar's
profile between 0.17 and 1.6~GHz have been extensively described in Wu
et al. (1998)\nocite{wgr+98}.
At the very lowest frequencies the pulsar has a clear
core component with two conal outriders. At higher frequencies the
pulse shape is much more amorphous and Wu et al. (1998) argue that it
has five blended components.  The pulse profile shows rather little
evolution with frequency above 1.4~GHz. The small amount of
polarization seen at this frequency has almost entirely
disappeared at 8.4~GHz.  There is a significant narrowing of the
profile, with the pulse width at 8.4~GHz being about half that at
1.4~GHz.  Following Wu et al. (1998) therefore, it seems as if the
outer conal components (which have a steep spectral index) have now
disappeared at 8.4~GHz entirely. It is also possible that the core
emission is now much weaker at this frequency, and the resultant
(narrow) profile is a blend of the two inner conal components.

\noindent
{\bf PSR J1522$-$5829 (B1518$-$58):} The 1.4~GHz profile shows a
simple Gaussian which likely consists of two equal strength components
as seen in the linear polarization profile \citep{qmlg95}. At
3.1~GHz the leading component is starting to dominate the profile and
this component is more highly polarized than its counterpart at
1.4~GHz (KJM05).  At 8.4~GHz the leading component now dominates the
profile, still has a high degree of linear polarization and the same
flat PA swing. It is difficult to tell whether this is simply a leading
edge conal profile or a symmetrical cone with a steep spectral index
on its trailing edge.

\noindent
{\bf PSR J1539$-$5626 (B1535$-$56):} At 1.4~GHz, the profile of the
pulsar appears to consist of a simple Gaussian \citep{qmlg95} albeit
at low time resolution.
However linear polarization is only seen on the trailing
half of the component, indicating that there is a blend of at least
two components.  At 3.1~GHz, three components are seen; a broad
leading component, a narrow central component and a highly polarized
narrow trailing component (KJM05). In our 8.4~GHz data, three
components are also present but the trailing component now dominates
and remains highly polarized. This pulsar therefore has a core component
flanked by conal outriders although again the outrides have very different
spectral index (and polarization) behaviour.
\begin{figure}
\resizebox{1.0\hsize}{!}{\includegraphics[angle=-90]{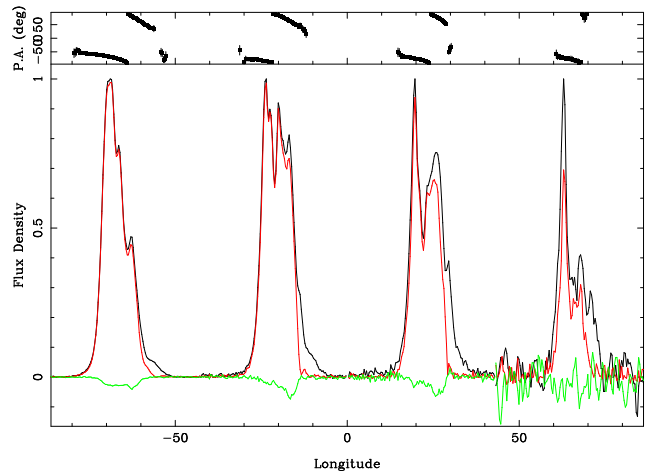}}
\caption{PSR~J0742$-$2822 at 0.66~GHz (extreme left), 1.4~GHz (middle left),
3.1~GHz (middle right) and 8.4~GHz (extreme right) in full polarization.
Note the dramatic changes in the total intensity profiles as a function
of frequency and the high linear polarization at all frequencies.}
\label{0742}
\end{figure}

\noindent
{\bf PSR J1600$-$5044 (B1557$-$50):} At frequencies below 1~GHz the
pulse profile is scatter broadened and not much structure can be discerned
\citep{vdhm97}. At 1.56~GHz the profile shows
two blended components with the trailing component showing moderate
linear polarization and very strong positive circular polarization
\citep{wmlq93}. At 3.1~GHz the two components are more clearly
split with the trailing component again having circular polarization in
excess of 50 per cent. The PA swing appears much steeper at 3.1~GHz than
at 1.6~GHz. In our 8.4~GHz observations the ratio of the
two peaks is similar to that at the lower frequencies. However both
the linear and circular polarization have significantly decreased.
This is a symmetric pulse profile with no core.
\begin{figure}
\resizebox{1.0\hsize}{!}{\includegraphics[angle=-90]{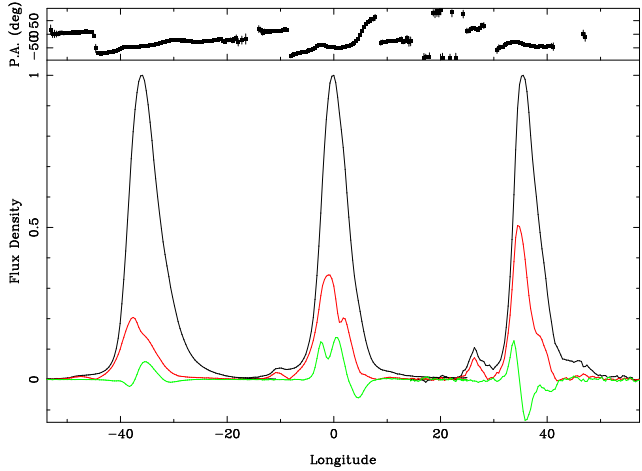}}
\caption{PSR~1644$-$4559 at 1.4~GHz (left), 3.1~GHz (middle) and
8.4~GHz (right) in full polarization. The pulse profile does not evolve
strongly with frequency although outriders are more prominent at the
highest frequency. Note the increasing linear polarization
and the peculiar evolution of the circular polarization.}
\label{1644}
\end{figure}

\noindent
{\bf PSR J1602$-$5100 (B1558$-$50):} The profile of this pulsar
undergoes significant frequency evolution. At 0.95~GHz there are two
components with the leading component dominating. At 1.4~GHz the profile 
is not greatly different (KJ06). A steep swing of PA over more than 200\degr\
is seen through the centre of the pulsar and there is also a swing
of circular polarization from negative to positive. At 3.1~GHz, the
ratio of the amplitudes of the components is reduced. This evolution appears
to accelerate because in our 8.4~GHz profile there remains only the merest
hint of the leading component and the trailing component completely dominates
the profile. This therefore appears to be a symmetrical double with
no core emission and an asymmetric spectral index for the two components.

\noindent
{\bf PSR J1630$-$4733 (B1627$-$47):} Scatter broadening dominates the
profile at low frequencies, but at 3.1~GHz the pulse shape is a simple
Gaussian, as is the circular polarization although there are two distinct
linear polarization features (KJM05). In our 8.4~GHz profile the pulse shape
is similar to that at 3.1~GHz although the linear polarization is
present only on the trailing half and the profile appears somewhat
narrower. This is therefore a blended symmetrical double.

\noindent
{\bf PSR J1644$-$4559 (B1641$-$45):} At 1.4~GHz the pulse profile
consists of 4 components with a small leading component and blend of
components within the main pulse. The circular polarization swings
from negative to positive near the pulse centre \citep{joh04}.  At
3.1~GHz the profile looks similar except that the small leading
component is starting to become more prominent.  However, the
fractional linear polarization has increased and the circular
polarization is very different. The PA variation with pulse longitude
is complex and different between the two frequencies (KJ06).  At
8.4~GHz the profile looks similar with the trailing component being
relatively brighter and a small trailing component starting to appear.
The linear fraction continues to increase at least on the leading part
of the pulse. The orthogonal jump in PA after the leading component is
at the same longitude at all frequencies.  The circular polarization
is again different from either of the two lower frequencies.
Figure~\ref{1644} shows the polarization profiles of the pulsar
at three frequencies. The profile is consistent with a leading edge cone.

\noindent
{\bf PSR J1709$-$4429 (B1706$-$44):} This pulsar is a young pulsar
with a characteristic simple profile, high degree of linear
polarization and a rather flat PA swing \citep{jw06}.
At 8.4~GHz its total intensity profile is virtually unchanged
relative to that a lower frequencies. The linear polarization remains
high and there is moderate circular polarization. It seems likely
that this is a grazing edge cone. This pulsar is the
brightest in our sample at 8.4~GHz with a flux density not
significantly different to that at 1.4~GHz.

\noindent
{\bf PSR J1721$-$3532 (B1718$-$35):} The profile of this pulsar
appears to be similar between 1.4 and 8.4~GHz (Qiao et al. 1995;
KJM05) although the lower frequency has a long scattering tail. There
is a slow rising edge to the profile followed by a steep falling edge.
Moderate linear and right-hand circular polarization is present at all
these frequencies. This is likely to be a trailing edge cone.

\noindent
{\bf PSR J1730$-$3350 (B1727$-$33):} The profile of this young pulsar at
8.4~GHz is highly linearly polarized despite the low signal to noise
ratio. Some right-hand circular polarization is also seen. A
comparison to an earlier observation at 1.4~GHz \citep{cmk01}
reveals little change in the fractional polarization between these two
frequencies.  The PA of the linear polarization is flat, both at 1.4
and 8.4~GHz. This is a typical young pulsar profile, likely a grazing edge cone.
\begin{table}
\caption{Pulsar classification. The three columns following the name
indicate whether the leading (l), central (c) and trailing (t) components
are polarized with f denoting flatter spectrum. Young pulsars are listed
first in their respective categories.}
\begin{tabular}{llcccc}
\hline & \vspace{-3mm} \\
\multicolumn{2}{c}{Profiles with cores} & l & c & t\\
\hline & \vspace{-3mm} \\
\multicolumn{2}{c}{Symmetric} \\
J0742$-$2822 & B0738$-$28 & $\bullet$f & $\bullet$ & $\bullet$ & young \\
J0837$-$4135 & B0835$-$41 & & & $\bullet$\\
J1243$-$6423 & B1240$-$64 & & & \\
J1326$-$5859 & B1323$-$58 & f & & \\
J1456$-$6843 & B1451$-$68 & & & \\
J1539$-$5626 & B1535$-$56 & & & $\bullet$f \\
\\
\multicolumn{2}{c}{Asymmetric} \\
J0835$-$4510 & B0833$-$45 &   & $\bullet$ & $\bullet$ & young \\
J1430$-$6623 & B1426$-$66 &   & & $\bullet$\\
J1453$-$6413 & B1449$-$64 & $\bullet$ & &   \\
J1752$-$2806 & B1749$-$28 &   & & \\
\hline & \vspace{-3mm} \\
\multicolumn{2}{c}{Profiles without cores} & l & c & t \\
\hline & \vspace{-3mm} \\
\multicolumn{2}{c}{Symmetric} \\
J0659+1414   & B0656+14   & &   & & young \\
J0908$-$4913 & B0906$-$49 & $\bullet$ &   & $\bullet$ & young \\
J1302$-$6350 & B1259$-$63 & $\bullet$ &   & $\bullet$ & young \\
J1341$-$6220 & B1338$-$62 & $\bullet$ &   & $\bullet$ & young \\
J0630$-$2834 & B0628$-$26 & &   & \\
J0738$-$4042 & B0736$-$40 & &   & \\
J1136+1551   & B1133+16   & f &   & \\
J1327$-$6222 & B1324$-$62 & f &   & \\
J1522$-$5829 & B1518$-$58 & $\bullet$f &   & \\
J1600$-$5044 & B1557$-$50 & &   & \\
J1602$-$5100 & B1558$-$50 & &   & f \\
J1630$-$4733 & B1627$-$47 & &   & \\
\\
\multicolumn{2}{c}{Asymmetric} \\
J1048$-$5832 & B1046$-$58 &   & $\bullet$ &   & young \\
J1709$-$4429 & B1706$-$44 &   & $\bullet$ &   & young \\
J1730$-$3350 & B1727$-$33 &   & $\bullet$ &   & young \\
J1740$-$3015 & B1737$-$30 &   & $\bullet$ &   & young \\
J0922+0638   & B0919+06   &   &   & $\bullet$\\
J0953+0755   & B0950+08   &   &   & \\
J1056$-$6258 & B1054$-$62 & $\bullet$ &   &   \\
J1359$-$6038 & B1356$-$60 & $\bullet$ &   &   \\
J1644$-$4559 & B1641$-$45 & $\bullet$ &   &   \\
J1721$-$3532 & B1718$-$35 &   &   & $\bullet$\\
\hline & \vspace{-3mm} \\
\end{tabular}
\end{table}
\noindent

\noindent
{\bf PSR J1740$-$3015 (B1737$-$30):} At all frequencies, the profile
is simple and the total polarization is high. The fractional linear
and circular polarization increases between 0.4 and 1.4~GHz
\citep{gl98}.  This pulsar retains a very high degree of polarization
at 4.8~GHz \citep{hkk98} and in our 8.4~GHz profile with 51\% linear
and 60\% circular polarization. The PA swing is the same at all
frequencies.  This pulsar is a virtual twin of PSR~B0144+59 which
shows a very similar evolution with frequency \citep{hkk98}.
Again, this is a typical young pulsar profile, likely a grazing edge cone.

\noindent
{\bf PSR J1752$-$2806 (B1749$-$28):} At frequencies below 1.4~GHz the
pulse profile looks very similar and consists of two blended
components (van Ommen et al. 1997, KJ06). At 3.1~GHz the profile has
significantly narrowed and the trailing component is strongly reduced
in amplitude (KJ06).  The PA swing is complex at different at 1.4~GHz
and 3.1~GHz.  Our 8.4~GHz observations show that the initial leading
component has continued to narrow. However, a strong trailing
component has now emerged at significantly later longitudes than the
component seen at low frequencies (and which may also be present in the
4.7~GHz observation of Sieber et al. 1975\nocite{srw75}). 
There is virtually no polarization
at this frequency. It seems likely that this is still a core dominated
profile with a trailing conal component.

\section{Discussion}
We can make several general observations about polarization at
8.4~GHz, reinforcing the conclusions come to by others. First, the
general fractional polarization is lower at high frequencies than at
low frequencies, apart from in the young pulsars.
Secondly, it is clear that conal emission generally
has a flatter spectral index than the core emission and hence the
cones become much more prominent at this frequency. Thirdly, the
overall profile width does not greatly decrease between 1.4 and
8.4~GHz; the general effect of width evolution is only really visible
below about 600~MHz.  One might expect that the PA swing at high
frequencies becomes simpler without the distorting influence of the
central component(s).  Unfortunately this inference is not obvious
because of the overall lack of polarization in the profiles.

Table~2 shows the classification of the pulsars in tabular form
according to the scheme laid out in Section 3. Three aspects can be
clearly identified:
\begin{enumerate}
\item The young pulsars all show a very high degree of polarization,
the only exception being PSR~J0659+1414 which abruptly depolarizes at
frequencies above about 4~GHz.
This result confirms the results found
by von Hoensbroech et al. (1998)\nocite{hkk98} who showed a correlation
between the polarized fraction and age. 
The polarized fraction remains approximately constant with frequency
although in some pulsars (e.g. Vela) the linear polarization decreases
while the circular polarization increases.
\item Core emission is generally lacking in young pulsars although
there are some exceptions. Johnston \& Weisberg (2006)\nocite{jw06}
comment extensively on the pulse morphology of young pulsars.
In those pulsars which do show core emission at 8.4~GHz,
virtually all lack polarization.
\item The pulsars with asymmetric cones tend to be relatively highly
polarized whether they are leading or trailing edge.  In contrast, the
symmetrical profiles have virtually no polarization. This is a rather
surprising result, and lends itself to no obvious explanation.
\end{enumerate}

In young pulsars the emission height is large even at high frequencies
\citep{jw06} and the emission remains polarized. In contrast core
emission in older pulsars arises from low in the magnetosphere
\citep{mr01} and quickly becomes less polarized with
frequency. Pulsars with a symmetrical conal configuration, likely to
be true conal rings, also appear to have low emission heights
\citep{gg03} and are also observed to show little or no polarization
at high frequencies. We therefore surmise that the fractional
polarization is determined by the emission height. Perhaps then the
asymmetric profiles are more symbolic of the patchy beam model
\citep{lm88} with conditions necessary for the production of radio
emission occurring higher in the magnetosphere.

In KJM05 we outlined a simple model whereby the various evolutionary
features of polarized components could be explained in the context of
the spectral index behaviour of the competing orthogonal modes. Three
types of behaviour are expected. In the first, the polarization is
high at all but very low frequencies and the spectral index is
shallow, in the second the polarization declines as a function of
frequency up to at least 10~GHz and the spectral index is steep. In
the third, the polarization fraction declines and then increases again
with a minimum in the GHz observing bands and one might also expect a
spectral break in the data.  All three of these types are seen in the
8.4~GHz profiles. The young pulsars which are highly polarized at low
frequencies remain so at high frequencies and their spectral indices
are reasonably flat (type 1 behaviour).  Furthermore, individual
highly polarized components such as that seen on the trailing edge of
PSR~J1539$-$5625 remain highly polarized and begin to dominate the
profile at high frequencies because of their flat spectral index.
Type 2 behaviour is most common, with declining polarization seen in
the majority of the non-young pulsars. Finally type 3 behaviour is
clearly seen in PSR~J1644$-$4559 with the polarization fraction
increasing between 1.4 and 8.4~GHz (see fig~\ref{1644}).

Highly polarized, type 1 components can account for the profiles of
young pulsars. However, highly polarized components are also 
often found in conjunction with other, less polarized components 
within individual profiles. These are
extremely interesting cases, in that the highly polarized component
may originate from higher in the magnetosphere than the rest of the
profile. Type 2 components are compatible with the standard model,
where higher frequencies originate from lower heights and therefore
are also less polarized. Type 3 are more complicated to explain in the
context of emission heights. Two possibilities exist. Either the
higher frequencies originate from low emission heights as the standard
model suggests, and the high polarization is set by some other
mechanism, or the components that are more polarized with frequency
originate further from the pulsar surface,
contrary to the standard behaviour. In this context it would be
useful to devise a technique to measure the emission heights of 
such components in an independent way (e.g. Gangadhara 2005\nocite{gan05}).
The tentative conclusion about the fractional
polarization and emission height that we have reached here could be
strengthened or challenged by theoretical considerations. It is
certainly tempting to start drawing conclusions suggesting a different
altitude of emission of various components within a single
profile. The impact of this on observational parameters such as the
component widths and the PA may provide interpretation to the most
complex observations \citep{kj06}, which do not adhere to the standard
pulsar model.

A physical model in a series of papers by Petrova (2001, 2002,
2003)\nocite{pet01,pet02,pet03} attempts to explain the observed
features of pulsar polarization in terms of orthogonal mode conversion
and refraction in the magnetosphere. In her model, a single emission
mode is produced which is later converted into two orthogonal
modes. This conversion best occurs in regions of longitudinal
propagation with respect to the magnetic field. In turn, these regions
occur where refraction is strongest. At high frequencies (or low
emission heights more strictly) refraction is effective, and therefore
mode conversion depolarizes the observed emission. Again, the high
altitude of the emission from young pulsars shields them from this
effect and the little mode conversion happens. This is also supported
by the relatively simple swing of the PA traverse in young pulsars.

\section{Conclusions}
We have substantially increased the number of pulsars with high
frequency data by producing calibrated polarization profiles for 32
objects.  Many of the features seen and the evolution of the profiles
from low to high frequency are generally as expected in the standard
picture of the observational phenomenology. Of most interest is the
continued high polarization fraction seen in the young pulsar profiles
and the curious result that asymmetric conal features are more highly
polarized than the symmetric features. The observations point towards
the fractional polarization being related to the emission height, with
polarized components originating from higher in the magnetosphere. It
is heartening that the recent theoretical models of Petrova and others
go some way towards explaining the diversity of features seen.

\section*{Acknowledgments}
The Australia Telescope is funded by the Commonwealth of 
Australia for operation as a National Facility managed by the CSIRO.
KW was supported by U.S. NSF Grant AST 0406832.

\label{lastpage}
\bibliographystyle{mn2e}
\bibliography{journals,modrefs,psrrefs,crossrefs}

\begin{thebibliography}{}

\bibitem[\protect\citeauthoryear{{Crawford}, {Manchester} \&
  {Kaspi}}{{Crawford} et~al.}{2001}]{cmk01}
{Crawford} F.,  {Manchester} R.~N.,    {Kaspi} V.~M.,  2001, AJ, 122, 2001

\bibitem[\protect\citeauthoryear{{Everett} \& {Weisberg}}{{Everett} \&
  {Weisberg}}{2001}]{ew01}
{Everett} J.~E.,  {Weisberg} J.~M.,  2001, ApJ, 553, 341

\bibitem[\protect\citeauthoryear{Gangadhara}{Gangadhara}{2005}]{gan05}
Gangadhara~R.~T., 2005, ApJ, 628, 923

\bibitem[\protect\citeauthoryear{Gould \& Lyne}{Gould \& Lyne}{1998}]{gl98}
Gould D.~M.,  Lyne A.~G.,  1998, MNRAS, 301, 235

\bibitem[\protect\citeauthoryear{{Gupta} \& {Gangadhara}}{{Gupta} \&
  {Gangadhara}}{2003}]{gg03}
{Gupta} Y.,  {Gangadhara} R.~T.,  2003, ApJ, 584, 418

\bibitem[\protect\citeauthoryear{Hamilton, McCulloch, Ables \&
  Komesaroff}{Hamilton et~al.}{1977}]{hmak77}
Hamilton P.~A.,  McCulloch P.~M.,  Ables J.~G.,    Komesaroff M.~M.,  1977,
  MNRAS, 180, 1

\bibitem[\protect\citeauthoryear{{Hotan}, {van Straten} \&
  {Manchester}}{{Hotan} et~al.}{2004}]{hvm04}
{Hotan} A.~W.,  {van Straten} W.,    {Manchester} R.~N.,  2004, PASA, 21, 302

\bibitem[\protect\citeauthoryear{{Johnston}}{{Johnston}}{2004}]{joh04}
{Johnston} S.,  2004, MNRAS, 348, 1229

\bibitem[\protect\citeauthoryear{Johnston, Hobbs, Vigeland, Kramer, Weisberg \&
  Lyne}{Johnston et~al.}{2005}]{jhv+05}
Johnston S.,  Hobbs G.,  Vigeland S.,  Kramer M.,  Weisberg J.~M.,    Lyne
  A.~G.,  2005, MNRAS, 364, 1397

\bibitem[\protect\citeauthoryear{{Johnston}, {van Straten}, {Kramer} \&
  {Bailes}}{{Johnston} et~al.}{2001}]{jvkb01}
{Johnston} S.,  {van Straten} W.,  {Kramer} M.,    {Bailes} M.,  2001, ApJ,
  549, L101

\bibitem[\protect\citeauthoryear{Johnston \& Weisberg}{Johnston \&
  Weisberg}{2006}]{jw06}
Johnston S.,  Weisberg J.~M.,  2006, MNRAS, In Press

\bibitem[\protect\citeauthoryear{{Karastergiou} \& {Johnston}}{{Karastergiou}
  \& {Johnston}}{2006}]{kj06}
{Karastergiou} A.,  {Johnston} S.,  2006, MNRAS, 365, 353 (KJ06)

\bibitem[\protect\citeauthoryear{{Karastergiou}, {Johnston} \&
  {Manchester}}{{Karastergiou} et~al.}{2005}]{kjm05}
{Karastergiou} A.,  {Johnston} S.,    {Manchester} R.~N.,  2005, MNRAS, 359,
  481 (KJM05)

\bibitem[\protect\citeauthoryear{Karastergiou, Johnston, Mitra, van Leeuwen \&
  Edwards}{Karastergiou et~al.}{2003}]{kjm+03}
Karastergiou A.,  Johnston S.,  Mitra D.,  van Leeuwen A.~G.~J.,    Edwards
  R.~T.,  2003, MNRAS, 344, L69

\bibitem[\protect\citeauthoryear{{Karastergiou}, {Kramer}, {Johnston}, {Lyne},
  {Bhat} \& {Gupta}}{{Karastergiou} et~al.}{2002}]{kkj+02}
{Karastergiou} A.,  {Kramer} M.,  {Johnston} S.,  {Lyne} A.~G.,  {Bhat}
  N.~D.~R.,    {Gupta} Y.,  2002, A\&A, 391, 247

\bibitem[\protect\citeauthoryear{Karastergiou, {von Hoensbroech}, Kramer,
  Lorimer, Lyne, Doroshenko, Jessner, Jordan \& Wielebinski}{Karastergiou
  et~al.}{2001}]{khk+01}
Karastergiou A.,  {von Hoensbroech} A.,  Kramer M.,  Lorimer D.,  Lyne A.,
  Doroshenko O.,  Jessner A.,  Jordan A.,    Wielebinski R.,  2001, A\&A, 379,
  270

\bibitem[\protect\citeauthoryear{Kramer}{Kramer}{1994}]{kra94}
Kramer M.,  1994, A\&AS, 107, 527

\bibitem[\protect\citeauthoryear{{Kramer}, {Karastergiou}, {Gupta}, {Johnston},
  {Bhat} \& {Lyne}}{{Kramer} et~al.}{2003}]{kkg+03}
{Kramer} M.,  {Karastergiou} A.,  {Gupta} Y.,  {Johnston} S.,  {Bhat} N.~D.~R.,
     {Lyne} A.~G.,  2003, A\&A, 407, 655

\bibitem[\protect\citeauthoryear{Kramer, Wielebinski, Jessner, Gil \&
  Seiradakis}{Kramer et~al.}{1994}]{kwj+94}
Kramer M.,  Wielebinski R.,  Jessner A.,  Gil J.~A.,    Seiradakis J.~H.,
  1994, A\&AS, 107, 515

\bibitem[\protect\citeauthoryear{Lyne \& Manchester}{Lyne \&
  Manchester}{1988}]{lm88}
Lyne A.~G.,  Manchester R.~N.,  1988, MNRAS, 234, 477

\bibitem[\protect\citeauthoryear{McCulloch, Hamilton, Manchester \&
  Ables}{McCulloch et~al.}{1978}]{mhma78}
McCulloch P.~M.,  Hamilton P.~A.,  Manchester R.~N.,    Ables J.~G.,  1978,
  MNRAS, 183, 645

\bibitem[\protect\citeauthoryear{Manchester}{Manchester}{1996}]{man96}
Manchester R.~N.,  1996, in Johnston S.,  Walker M.~A.,   Bailes M.,  eds,
  Pulsars: Problems and Progress, {IAU} Colloquium 160.
Astronomical Society of the Pacific, San Francisco, pp 193--196

\bibitem[\protect\citeauthoryear{Manchester \& Johnston}{Manchester \&
  Johnston}{1995}]{mj95}
Manchester R.~N.,  Johnston S.,  1995, ApJ, 441, L65

\bibitem[\protect\citeauthoryear{Mitra \& Rankin}{Mitra \& Rankin}{2002}]{mr01}
Mitra D.,  Rankin J.~M.,  2002, ApJ, 577, 322

\bibitem[\protect\citeauthoryear{Morris, Graham, Seiber, Bartel \&
  Thomasson}{Morris et~al.}{1981}]{mgs+81}
Morris D.,  Graham D.~A.,  Seiber W.,  Bartel N.,    Thomasson P.,  1981,
  A\&AS, 46, 421

\bibitem[\protect\citeauthoryear{{Petrova}}{{Petrova}}{2001}]{pet01}
{Petrova} S.~A.,  2001, A\&A, 378, 883

\bibitem[\protect\citeauthoryear{{Petrova}}{{Petrova}}{2002}]{pet02}
{Petrova} S.~A.,  2002, A\&A, 383, 1067

\bibitem[\protect\citeauthoryear{{Petrova}}{{Petrova}}{2003}]{pet03}
{Petrova} S.~A.,  2003, A\&A, 408, 1057

\bibitem[\protect\citeauthoryear{Phillips \& Wolszczan}{Phillips \&
  Wolszczan}{1992}]{pw92}
Phillips J.~A.,  Wolszczan A.,  1992, ApJ, 385, 273

\bibitem[\protect\citeauthoryear{Qiao, Manchester, Lyne \& Gould}{Qiao
  et~al.}{1995}]{qmlg95}
Qiao G.~J.,  Manchester R.~N.,  Lyne A.~G.,    Gould D.~M.,  1995, MNRAS, 274,
  572

\bibitem[\protect\citeauthoryear{Rankin}{Rankin}{1983}]{ran83}
Rankin J.~M.,  1983, ApJ, 274, 333

\bibitem[\protect\citeauthoryear{Rankin}{Rankin}{1990}]{ran90}
Rankin J.~M.,  1990, ApJ, 352, 247

\bibitem[\protect\citeauthoryear{Sieber}{Sieber}{1997}]{sie97}
Sieber W.,  1997, A\&A, 321, 519

\bibitem[\protect\citeauthoryear{Sieber, Reinecke \& Wielebinski}{Sieber
  et~al.}{1975}]{srw75}
Sieber W.,  Reinecke R.,    Wielebinski R.,  1975, A\&A, 38, 169

\bibitem[\protect\citeauthoryear{van Ommen, D'Alesssandro, Hamilton \&
  McCulloch}{van Ommen et~al.}{1997}]{vdhm97}
van Ommen T.~D.,  D'Alesssandro F.~D.,  Hamilton P.~A.,    McCulloch P.~M.,
  1997, MNRAS, 287, 307

\bibitem[\protect\citeauthoryear{von Hoensbroech}{von
  Hoensbroech}{1999}]{hoe99}
von Hoensbroech A.,  1999, PhD thesis, University of Bonn

\bibitem[\protect\citeauthoryear{von Hoensbroech, Kijak \& Krawczyk}{von
  Hoensbroech et~al.}{1998}]{hkk98}
von Hoensbroech A.,  Kijak J.,    Krawczyk A.,  1998, A\&A, 334, 571

\bibitem[\protect\citeauthoryear{von Hoensbroech \& Lesch}{von Hoensbroech \&
  Lesch}{1999}]{hl99}
von Hoensbroech A.,  Lesch H.,  1999, A\&A, 342, L57

\bibitem[\protect\citeauthoryear{von Hoensbroech \& Xilouris}{von Hoensbroech
  \& Xilouris}{1997a}]{hx97a}
von Hoensbroech A.,  Xilouris K.~M.,  1997a, A\&A, 324, 981

\bibitem[\protect\citeauthoryear{von Hoensbroech \& Xilouris}{von Hoensbroech
  \& Xilouris}{1997b}]{hx97}
von Hoensbroech A.,  Xilouris K.~M.,  1997b, A\&AS, 126, 121

\bibitem[\protect\citeauthoryear{{Wang}, {Johnston} \& {Manchester}}{{Wang}
  et~al.}{2004}]{wjm04}
{Wang} N.,  {Johnston} S.,    {Manchester} R.~N.,  2004, MNRAS, 351, 599

\bibitem[\protect\citeauthoryear{{Weisberg}, {Cordes}, {Kuan}, {Devine},
  {Green} \& {Backer}}{{Weisberg} et~al.}{2004}]{wck+04}
{Weisberg} J.~M.,  {Cordes} J.~M.,  {Kuan} B.,  {Devine} K.~E.,  {Green} J.~T.,
     {Backer} D.~C.,  2004, ApJS, 150, 317

\bibitem[\protect\citeauthoryear{Weisberg, Cordes, Lundgren, Dawson, Despotes,
  Morgan, Weitz, Zink \& Backer}{Weisberg et~al.}{1999}]{wcl+99}
Weisberg J.~M.,  Cordes J.~M.,  Lundgren S.~C.,  Dawson B.~R.,  Despotes J.~T.,
   Morgan J.~J.,  Weitz K.~A.,  Zink E.~C.,    Backer D.~C.,  1999, ApJS, 121,
  171

\bibitem[\protect\citeauthoryear{{Wu}, {Gao}, {Rankin}, {Xu} \&
  {Malofeev}}{{Wu} et~al.}{1998}]{wgr+98}
{Wu} X.,  {Gao} X.,  {Rankin} J.~M.,  {Xu} W.,    {Malofeev} V.~M.,  1998, AJ,
  116, 1984

\bibitem[\protect\citeauthoryear{Wu, Manchester, Lyne \& Qiao}{Wu
  et~al.}{1993}]{wmlq93}
Wu X.,  Manchester R.~N.,  Lyne A.~G.,    Qiao G.,  1993, MNRAS, 261, 630

\bibitem[\protect\citeauthoryear{Xilouris, Kramer, Jessner, Wielebinski \&
  Timofeev}{Xilouris et~al.}{1996}]{xkj+96}
Xilouris K.~M.,  Kramer M.,  Jessner A.,  Wielebinski R.,    Timofeev M.,
  1996, A\&A, 309, 481

\bibitem[\protect\citeauthoryear{Xilouris, Seiradakis, Gil, Sieber \&
  Wielebinski}{Xilouris et~al.}{1995}]{xsg+95}
Xilouris K.~M.,  Seiradakis J.~H.,  Gil J.~A.,  Sieber W.,    Wielebinski R.,
  1995, A\&A, 293, 153

\end{thebibliography}
\end{document}